\documentclass[11pt]{article}
\usepackage{authblk}
\usepackage{multirow}
\usepackage{xcolor}
\usepackage{setspace}
\usepackage{amsmath}
\usepackage{mathtools}
\usepackage{bm}
\usepackage{amsthm}
\usepackage{amssymb}
\usepackage[ruled]{algorithm2e}
\usepackage{graphicx,psfrag,epsf}
\usepackage{enumerate}
\usepackage{subcaption}
\usepackage{caption}
\usepackage{natbib}
\usepackage{tikz}
\usepackage{url}  
\usepackage{appendix}
\usepackage{titlesec}
\usepackage{etoolbox}
\usepackage{makecell}
\usepackage{booktabs}
\usepackage{geometry}
\makeatletter
\def\@seccntformat#1{\@ifundefined{#1@cntformat}%
   {\csname the#1\endcsname\quad}
   {\csname #1@cntformat\endcsname}}
\apptocmd\appendix{%
    \newcommand\section@cntformat{\appendixname\ }
    \addtocontents{toc}{\bigskip\noindent\textbf{Appendix Material}\par}
    {}{}}
\makeatother

\newcommand{\blind}{1}

\newtheorem{remark}{Remark}

\begin{document}

\def\spacingset#1{\renewcommand{\baselinestretch}%
{#1}\small\normalsize} \spacingset{1}


\if1\blind
{
  \title{\bf RECaST-Surv: A Calibrated Borrowing Method for Survival Endpoints in Unequal Randomized Trials} 
  \author[1]{Dehua Bi}  
  \author[1]{Arlina Shen}
  \author[2]{Ruben P.A. van Eijk}
  \author[1]{Lu Tian}
  \author[3]{Jiapeng Xu}
  \author[4]{Guillemette de la Borderie}
  \author[4]{Nate Bennet}
  \author[4]{Sarno Maria}
  \author[1]{Ying Lu}

  \affil[1]{Department of Biomedical Data Science, Stanford University, CA}
  \affil[2]{Department of Neurology, UMC Utrecht Brain Center, University Medical Center Utrecht, Utrecht, The Netherlands}
  \affil[3]{University of Cambridge, Cambridge, UK}
  \affil[4]{UCB Pharma, Brussels, Belgium}
  \maketitle
} \fi

\if0\blind
{
  \bigskip
  \bigskip
  \bigskip
  \begin{center}
    {\LARGE\bf RECaST-Surv: A Calibrated Borrowing Method for Survival Endpoints in Unequal Randomized Trials}
\end{center}
  \medskip
} \fi

\bigskip
\begin{abstract}

Randomized trials with limited concurrent control information---including but not limited to unequal-randomization settings---can offer ethical and practical advantages, especially in pediatric and rare diseases, but they often lose power because fewer control patients are available for direct comparison. Borrowing information from external controls may improve efficiency, but can also inflate the Type I error rate when the external and trial populations are not sufficiently comparable. We propose RECaST-Surv, a Bayesian transfer-learning framework for time-to-event outcomes that extends the RECaST method to survival settings. The method learns a structural survival model from external control data and calibrates it to the concurrent control arm of the current trial through a Cauchy random effect. To improve frequentist operating characteristics, we further develop a bootstrap-based procedure to calibrate the testing rule for Type I error control. RECaST-Surv can accommodate multiple external datasets and requires only summary-level information from external sources. Simulation studies show that the method maintains near-nominal Type I error across challenging settings while improving power by roughly 10\%--12\% over standard analyses. In an Amyotrophic Lateral Sclerosis trial emulation, RECaST-Surv increased power from 82.8\% to 95.7\% relative to the no-borrowing RCT analysis, while maintaining acceptable error control.

\end{abstract}

\noindent


\vfill

\newpage
\spacingset{1.45}

\section{Introduction}\label{sec:intro}

Randomized controlled trials (RCTs) are the gold standard for evaluating the effect of a new treatment relative to a control. Depending on the development stage, trial objectives differ. For example, Phase I dose-finding trials are primarily designed to assess toxicity and identify the maximum tolerated dose (MTD). Later-phase trials may aim to demonstrate noninferiority to the current standard of care or, when no effective treatment exists, superiority over placebo. In either case, well-designed randomized trials provide the most reliable evidence for evaluating treatment effects.

Equal randomization in RCTs is common in practice due to its ease of implementation and statistical efficiency for a fixed total sample size. Nevertheless, enrolling patients under equal randomization can sometimes be challenging or less desirable, especially in rare diseases, pediatric trials, or settings where an $r:1$ randomization ratio is used to enhance patient enrollment. In these settings, investigators may prefer to assign more patients to the experimental arm, i.e., unequal randomization. Although unequal randomization generally requires a larger total sample size to achieve the same power as equal allocation, it may offer ethical and practical advantages. For example, it can increase the chance of receiving the experimental treatment, improve enrollment, and accommodate settings where differential dropout between arms is anticipated. As a result, unequal randomization has become increasingly common in recent years, particularly in oncology \citep{gupta2021unequal, nay2024justification}. More generally, the challenge addressed in this paper is not limited to unequal randomization. It also arises in some $1:1$ randomized trials when the overall trial size is small or when the concurrent control arm is otherwise limited.


At the same time, the growing availability of real-world data (RWD), including historical trial data and electronic health records (EHRs), has motivated the development of statistical methods for borrowing information from external sources. These methods aim to augment the control arm of the current trial and construct a hybrid control, thereby improving efficiency and increasing power when the concurrent control sample size is limited. A large class of these methods borrows information primarily through outcome models. Bayesian methods, including the power prior (PP) \citep{ibrahim2000power}, modified power prior \citep{duan2006using}, commensurate prior (CP) \citep{hobbs2012commensurate}, meta-analytic-predictive (MAP) prior \citep{neuenschwander2010summarizing}, and robust MAP (R-MAP) prior \citep{schmidli2014robust}, are widely used for this purpose. More recent developments, such as the Bayesian Hybrid Design with Flexible Sample Size Adaptation (BEATS) \citep{bi2023beats}, further extend this line of work by enabling adaptive borrowing in more flexible trial settings. These outcome-based approaches typically rely, either explicitly or implicitly, on some form of exchangeability between the external data and the current-trial control arm; that is, after accounting for the modeled quantities, the external and current-trial controls are assumed to arise from sufficiently similar distributions that information can be shared without substantial bias.

Another line of work incorporates baseline covariates to relax this exchangeability assumption between the current trial and external data. Propensity score (PS)-based approaches, including PS-integrated composite likelihood (PSCL) \citep{chen2020propensity}, PS-PP \citep{lu2022propensity}, and PS-MAP \citep{liu2021propensity}, use matching or weighting strategies to selectively borrow from external patients who are similar to those in the current trial. Other methods employ more flexible modeling tools to identify exchangeable subgroups. For example, the Shared Atoms Model for Hybrid Control (SAM-HC) \citep{bi2023pam} constructs synthetic or hybrid controls by identifying common subpopulations across the trial and external data. The Latent Exchangeability Prior (LEAP) \citep{alt2023leap} enables dynamic borrowing from exchangeable subsets, while Bayesian Additive Regression Trees (BART) \citep{chipman2010bart} have also been used to integrate multiple data sources and adjust for covariates \citep{zhou2021incorporating}.

These methods broaden the ways in which information can be borrowed from external data and help relax the strict exchangeability assumptions required by outcome-based approaches. However, they do not eliminate the basic concern that the current trial and external data may still differ in ways not fully captured by observed covariates or modeled latent structure. Differences in patient populations, study conduct, or unmeasured confounding may therefore persist, and when this occurs, information borrowing can introduce bias and inflate the Type I error rate. Moreover, in most Bayesian borrowing approaches, frequentist operating characteristics, especially the Type I error rate, are usually examined retrospectively through simulation studies rather than addressed in a study-specific manner at the design or analysis stage. As a result, the impact of borrowing on Type I error is often evaluated under prespecified scenarios and may not fully reflect the features of the actual trial at hand.

Against this background, transfer learning provides an alternative framework for borrowing information from external sources. Unlike the borrowing methods described above, which aim to justify direct information sharing through exchangeability assumptions or covariate-based adjustment, transfer learning starts from the premise that the external and trial populations may differ in systematic ways. It uses the external data to learn a source model and then adapts that model to the target population, rather than treating the two populations as interchangeable. In this framework, the domain refers to the covariate space and its marginal distribution, whereas the task refers to the outcome being predicted and the associated prediction rule. Recently, \cite{hickey2024transfer} proposed the Random Effect Calibration of Source to Target (RECaST) method, which uses a Cauchy random effect to calibrate a pretrained source model to the target population. A practical advantage of RECaST is that it requires only summary-level information from the source model---specifically, the fitted coefficient vector that defines the source-model linear predictor---rather than patient-level source data. This feature is useful when external data are distributed across institutions and individual-level sharing is difficult because of privacy concerns and the administrative burden of data transfer.

In this work, we extend the RECaST framework to time-to-event outcomes and to randomized trials with limited concurrent control information, with unequal randomization as a primary motivating setting. More importantly, we augment the borrowing procedure with a bootstrap-based cross-validation step that provides a trial-specific way to address Type I error using only currently observed trial data. The calibration is performed with the RCT control arm and exploits randomization to approximate the null behavior of the analysis under the current study. In this sense, the proposed procedure separates information borrowing for estimation from calibration for hypothesis testing. Unlike approaches that rely on extensive prespecified simulations to study Type I error under hypothetical scenarios, our procedure uses observed trial data to define a study-specific calibration rule. We do not claim that this guarantees exact nominal Type I error in finite samples. Rather, the goal is to provide a practical, data-driven way to account for Type I error at the trial level under observed study conditions.

This distinguishes RECaST-Surv from existing Bayesian borrowing approaches, in which Type I error is usually investigated after analysis through simulation rather than incorporated through a calibration step tied to the current trial. Our contributions are threefold. First, we extend RECaST to survival outcomes. Second, we develop a bootstrap cross-validation procedure that provides a study-specific calibration for Type I error in trials with limited concurrent control information, including unequal randomized trials. Third, we generalize the framework to combine predictions from multiple external sources. Together, these developments yield a practical framework for borrowing external information in survival trials while explicitly considering frequentist operating characteristics at the trial level.

The remainder of the paper is organized as follows. Section \ref{sec:overview_RECaST} reviews the original RECaST framework. Section \ref{sec:method} presents the proposed RECaST-Surv method and describes how it is used for information borrowing. Section \ref{sec:sim} reports simulation studies, and Section \ref{sec:data_ana} presents a real-data application. We conclude with a discussion in Section \ref{sec:conclude}.

\section{Review of RECaST}\label{sec:overview_RECaST}

We briefly review the original RECaST framework introduced in \cite{hickey2024transfer}. Here, the terms \emph{source} and \emph{target} are used in the usual transfer-learning sense: the source dataset is a larger dataset used to build an initial model, and the target dataset is a smaller dataset to which that model is adapted. At this stage, these labels are only generic and should not yet be interpreted as the external and current-trial datasets in the hybrid-design setting studied in this paper; that specific connection is made later.

Let $j \in \{s,t\}$ index the source and target datasets, respectively. For dataset $j$, let $\mathcal{Y}_j = \{y_{j,i}; i = 1, \ldots, n_j\}$ denote the observed outcomes, and let $\mathcal{X}_j = \{\bm{x}_{j,i}; i = 1, \ldots, n_j\}$ denote the corresponding covariates. We assume that the same set of covariates is observed in both datasets and that these covariates have been centered and scaled. We further assume that the relationship between covariates and outcome in the target and source data can be described as
\begin{equation}
y_{t,i} = h\{g(\bm{\theta}_t, \bm{x}_{t,i}), U_{t,i}\}, \text{ and }
y_{s,i} = h\{f(\bm{\theta}_s, \bm{x}_{s,i}), U_{s,i}\},
\label{eq:generate_model}
\end{equation}
respectively, where $f(\cdot, \cdot)$ and $g(\cdot, \cdot)$ denote the structural components of the source and target models, respectively, $h(\cdot,\cdot)$ is a link function with a known parametric form, $U_{j,i}, \,\, j=t, s$, are auxiliary random variables with known distributions, and $\bm{\theta}_j, \,\, j = t, s$ denote the corresponding model parameters, such as regression coefficients (and an intercept term).


The key idea of RECaST is to use the source model as a baseline and then calibrate it to the target population. To this end, RECaST introduces a patient-specific calibration term
$$
\beta_i = \frac{g(\bm{\theta}_t,\bm{x}_{t,i})}{f(\bm{\theta}_s,\bm{x}_{t,i})},
$$
which measures the deviation between the target and source models at covariate value $\bm{x}_{t,i}$. When the source and target share similar structure, $\beta_i$ is expected to be close to 1. Therefore, even if $\bm{\theta}_t$ cannot be estimated reliably from the target data alone because of limited sample size, the source-model estimate $\hat{\bm{\theta}}_s$, obtained from a much larger sample, may still be used as a baseline to model the target outcome via a calibrated model:
\begin{equation}\label{eq:recast_y_t}
    y_{t,i} = h\{\beta_i \cdot f(\hat{\bm{\theta}}_s,\bm{x}_{t,i}), U_{t,i}\}.
\end{equation}
To model $\beta_i$, \cite{hickey2024transfer} considered a first-order approximation and showed that $\beta_i$
approximately follows a Cauchy distribution, i.e., 
$$
\beta_i \sim \text{Cauchy}(\mu,\tau).
$$ 
Accordinly, RECaST uses a  hierarchical Bayesian model with the patient-specific calibration term $\beta_i$ following a Cauchy distribution $\text{Cauchy}(\mu, \tau).$ 
In this model, appropriate priors are assigned to $\mu$ and $\tau$. 
The main use of this heirachical Bayesian model is to generate posterior predictive outcomes for a group of observations with covariates $\tilde{\mathcal{X}}_t=\{\tilde{\bm{x}}_{t, i}, i=1, \cdots, \tilde{n}\}$, denoted by $\tilde{\mathcal{Y}}_t = \{\tilde{y}_{t,i};\, i = 1, \ldots, \tilde{n}\}$, which can be used to augment the statistical inference in the target population. 


Specifically, to sample from $\tilde{y}_{t,i} \mid \mathcal{X}_t, \mathcal{Y}_t, \hat{\bm{\theta}}_s, \tilde{\bm{x}}_{t,i}$, one may proceed as follows:
\begin{itemize}
    \item[1)] draw posterior samples $\tilde{\mu}$ and $\tilde{\tau}$ from the MCMC output, 
    \item[2)] sample $\tilde{\beta}_i \sim \text{Cauchy}(\tilde{\mu}, \tilde{\tau})$ and $U_{t,i}$ from the auxilirary distribuiton, 
    \item[3)] compute
    $
    \tilde{y}_{t,i} = h\{\tilde{\beta}_i \cdot f(\hat{\bm{\theta}}_s,\bm{x}_{t,i}), U_{t,i}\}.
    $
\end{itemize}
See Algorithms 1 and 2 in \cite{hickey2024transfer} for additional details.

A practical advantage of RECaST is that it requires only estimated source-model parameters rather than patient-level source data. This feature avoids some of the logistical and legal difficulties that often arise when data must be shared across hospitals or institutions. In addition, the framework can generate posterior predictive values for target patients under alternative treatment assignments, which may be interpreted as counterfactual predictions or ``digital twins.'' These properties make RECaST appealing in settings where the current trial is small but relevant external data are available.

Although RECaST is a general transfer-learning framework, the original work by \cite{hickey2024transfer} focused on continuous and binary outcomes. In this paper, we extend RECaST to survival outcomes and focus on trials with limited concurrent control information, particularly unequal-randomization trials in which more patients are assigned to the treatment arm than to the control arm.

\section{Method}\label{sec:method}

\subsection{Overview}\label{subsec:ov}

Consider a randomized controlled trial (RCT) with an $r:1$ ($r>1$) randomization ratio between the treated and control arms. Our goal is to estimate the treatment effect, expressed through a hazard ratio, by borrowing information from $J$ external data sources. We assume that these external sources contain real-world observations from hospitals or institutions that use the same control regimen as the current trial.

With a slight abuse of notation, let $j \in \{0,1,\ldots,J\}$ index the datasets. Here, $j=0$ denotes the current trial, and $j=1,\ldots,J$ denote the external sources. Let $n_0$ be the total sample size of the current trial, with $n_{0,1}$ treated patients and $n_{0,0}$ control patients such that $n_{0,1}/n_{0,0}\approx r$. For each dataset $j$, let
\[
\mathcal{Y}_j=\{(t_{j,i},\delta_{j,i}); i=1,\ldots,n_j\}, \quad
\mathcal{X}_j=\{\bm{x}_{j,i}; i=1,\ldots,n_j\}, 
\]
denote the observed follow-up times and event indicators, the baseline covariates, and the treatment indicators, respectively.  Specifically, the observed follow-up $t_{j,i}=\min(y_{j,i}, c_{j,i}),$ where $y_{j,i}$ is the event time of interest and $c_{j,i}$ is the censoring time. The event indicator $\delta_{j,i}=I(y_{j,i}<c_{j,i})$ indicates whether event time is observed. 
We assume that the same set of covariates is observed across all datasets, and that these covariates have been centered and scaled. 
All patients from external studies are in the control group, while  
patients in the target trial are randomized into treatment and control groups. We separate data from target trial into treated group and control group accordingly:
$$\mathcal{Y}_0^{(k)}=\{(t_{0,i}^{(k)},\delta_{0,i}^{(k)}); i=1,\ldots,n_{0,k}\}, 
\quad
\mathcal{X}_0^{(k)}=\{\bm{x}_{0,i}^{(k)}; i=1,\ldots,n_{0,k}\}$$
where $k=1$ and 0 represents treatment and control groups, respectively.

To explain the basic idea, we first consider the case $J=1$ with $n_1 \gg n_{0,0}$. Because the control arm in the current trial is small, we treat the current-trial controls as the \textit{target} data and the external controls as the \textit{source} data, and fit the RECaST model using these two datasets. The fitted model is then used to predict the counterfactual control survival for the treated patients in the current trial based on their observed covariates. Since treatment assignment within the trial is randomized, this prediction provides a natural basis for estimating the treatment effect in the treated population.

\begin{figure}[ht]
    \centering
    \includegraphics[width=\textwidth]{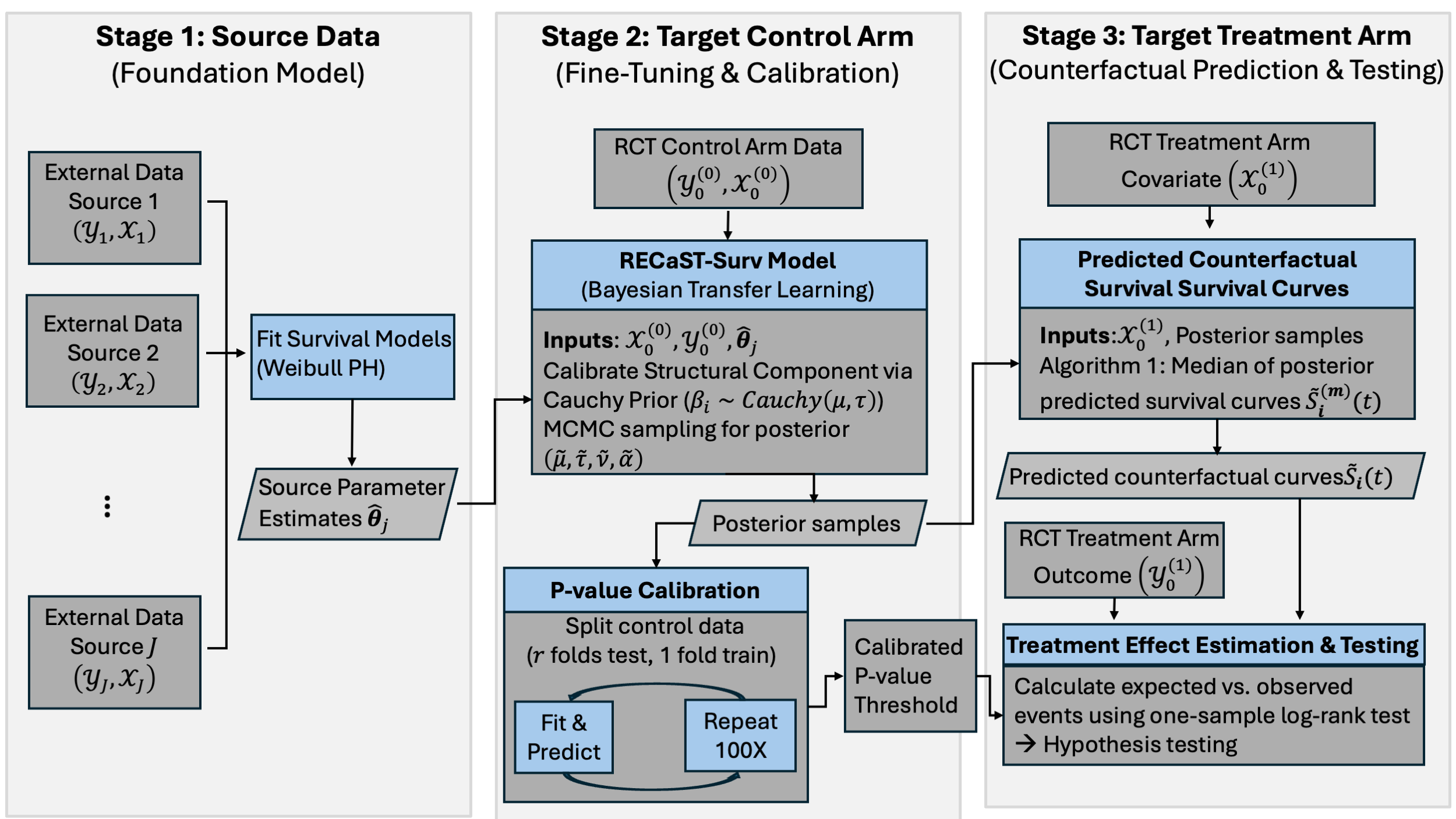}
    \caption{Schematic overview of the RECaST-Surv framework. The procedure has three stages: (1) estimation of source-model parameters from external data; (2) Bayesian transfer learning using the current-trial control arm together with bootstrap-based calibration; and (3) prediction of counterfactual control survival for the treated patients to estimate the treatment effect and conduct hypothesis testing.}
    \label{fig:flowchart}
\end{figure}

The proposed method proceeds in three steps. First, we fit a survival version of RECaST using the external source and the current-trial control arm. Second, we use the fitted model to generate counterfactual control survival curves for the treated patients and construct a test statistic for the treatment effect. Third, we calibrate the rejection rule using only the observed control-arm data through repeated splits that mimic the trial's $r:1$ randomization ratio. Figure \ref{fig:flowchart} provides a schematic summary.

The original RECaST framework in \cite{hickey2024transfer} was developed for continuous and binary outcomes. In what follows, we first extend it to time-to-event outcomes, which we call RECaST-Surv. We then describe how the resulting model is used in trials with limited concurrent control information, using unequal randomization as the primary motivating design, and finally extend the design to the case of multiple external sources.

\subsection{RECaST-Surv: RECaST for Survival Outcomes}

We extend RECaST to time-to-event data through a Weibull proportional hazards model. This gives a parametric survival model that is compatible with the generative structure of RECaST.

For simplicity, suppose $J=1$. Let $\hat{\bm{\theta}}_1$ denote the parameter estimate obtained from the external data $(\mathcal{Y}_1, \mathcal{X}_1)$, for example from a proportional hazards model fit. 
In the current implementation, the summary-level source information required from external source 1 is the fitted coefficient vector $\hat{\bm{\theta}}_1$ for the baseline covariates. Thus, once $\hat{\bm{\theta}}_1$ has been obtained, patient-level source data are no longer needed in the RECaST-Surv fitting step. In practice, this requires that the same set of baseline covariates are observed in the source and target datasets and that their definitions, centering, and scaling are harmonized across datasets.

Under RECaST-Surv, the hazard function for a control patient from the target study is modeled as
\[
\lambda(t \mid \beta,\hat{\bm{\theta}}_1,\bm{x}_{0})
=
\nu t^{\nu-1}\exp\{\alpha+\beta f(\bm{\theta}_1,\bm{x}_{0})\},
\]
where $\bm{x}_0$ is the baseline covariate, $\beta$ is the patient specific calibration factor, $\nu>0$ is the Weibull shape parameter, $\alpha$ is an intercept term,
and the source structural component 
\[
f(\bm{\theta}_1,\bm{x})=\bm{x}^T\bm{\theta}_1.
\]
 The corresponding survival function is
\begin{equation}
S(t \mid \beta,\hat{\bm{\theta}}_1,\bm{x}_{0})
=\exp\left[-t^\nu \exp\left\{\alpha+\beta f(\hat{\bm{\theta}}_1, \bm{x}_{0})\right\}\right]. 
 \label{eq:surv_func}
\end{equation}

Therefore, given $\beta$ and $(\alpha, \nu)$,  the event time can be generated from
\[
y_{0}
=
\left[
\frac{U_{0}}
{\exp\{\alpha+\beta f(\hat{\bm{\theta}}_1, \bm{x}_0)\}}
\right]^{1/\nu}.
\]
where $U_{0}$ follows an unit exponential distribution. 





A key step of RECaST-surv is to generate the survival function of a control patient with covariate $\bm{x}_0$ in the target study, denoted by $S(t\mid \bm{x}_0)= P\left(y_0^{(0)}>t\mid \bm{x}_0^{(0)}\right).$  Such survival functions can be used to sample synthetic outcomes and used for downstream inference. 

To this ends, we assume
\begin{enumerate}
\item $ \beta \sim \text{Cauchy}(\mu,\tau)$ as in the orignal RECaST framework,
\item appropriate priors for $(\mu,\tau,\nu,\alpha)$: 
independent Log-Normal priors for $(\tau,\nu)$ and independent Normal priors for $(\mu,\alpha)$ considering their range constraint, i.e., 
$$\mu\sim N(0, \sigma_\mu^2), \alpha\sim N(0, \sigma_\alpha^2), \tau \sim \exp\left\{ N(\mu_\tau, \sigma_\tau^2)\right\}\mbox{ and } \nu\sim \exp\left\{N(\mu_\nu, \sigma_\nu^2)\right\}.$$
\end{enumerate}

With those preparations, we may circumvent the need of directly sampling patient-specific calibration effects $\beta_i$ by sampling $(\mu, \tau, \nu, \alpha)$ from their posterior distributions. First, the marginal likelihood for observation $i$ in the control group of the target study is
\begin{equation}
\bar{L}\left(\bm{\psi}\mid\hat{\bm{\theta}}_1, t_{0,i}^{(0)}, \delta_{0,i}^{(0)}, \bm{x}_{0,i}^{(0)} \right)
=
\int
L\left( \beta, \bm{\psi} \mid \hat{\bm{\theta}}_1, t_{0,i}^{(0)}, \delta_{0,i}^{(0)}, \bm{x}_{0,i}^{(0)}\right)
\, \pi(\beta\mid \mu, \tau)\, d\beta,
\label{eq:marginal-lik}
\end{equation}
where $\bm{\psi}=(\mu,\tau,\nu,\alpha),$ 
$\pi(\cdot\mid \mu, \tau)$ is the density function for Cauchy$(\mu, \tau)$ and 
\[
L\left(\beta, \bm{\psi}\mid \hat{\bm{\theta}}_1, t_{0,i}^{(0)}, \delta_{0, i}^{(0)}, \bm{x}_{0,i}^{(0)}\right)
=
\left\{
\lambda\left(t_{0,i}^{(0)}\mid \beta,\hat{\bm{\theta}}_1,\bm{x}_{0,i}^{(0)}\right)
\right\}^{\delta_{0,i}^{(0)}}
S\left(t_{0,i}^{(0)}\mid \beta,\hat{\bm{\theta}}_1,\bm{x}_{0,i}^{(0)}\right).
\]
The final posterior distribution of $\bm{\psi}$ given all observations in the control group of the target study is 
\[
\pi\left(\bm{\psi}\mid {\cal Y}_0^{(0)}, {\cal X}_0^{(0)}\right)
\propto
\left\{
\prod_{i=1}^{n_{0,0}} \bar{L}\left(\bm{\psi}\mid\hat{\bm{\theta}}_1, t_{0,i}^{(0)}, \delta_{0,i}^{(0)}, \bm{x}_{0,i}^{(0)} \right)\right\}
\pi_\mu(\mu)\pi_\tau(\tau)\pi_\nu(\nu)\pi_\alpha(\alpha),
\]
where $\pi_\mu(\cdot), \pi_\tau(\cdot), \pi_\nu(\cdot)$ and $\pi_\alpha(\cdot)$ are density functions of the prior distributions for $\mu, \tau, \nu,$ and $\alpha,$ respectively. 

We propose to sample $\psi$
using a random-walk Metropolis--Hastings algorithm. At iteration $m$, we propose sample
\[
\bm{\phi}^* \sim MVN\left(\bm{\phi}^{(m-1)},\, s_m\Sigma_m\right),
\]
and let
$$ \bm{\psi}^{(m)}=\begin{cases}  \bm{\psi}^* &\mbox{ with probability }~~ \pi\\                                   
\bm{\psi}^{(m-1)} &\mbox{ with probabilty}~~ 1-\pi
\end{cases},
$$
where  $s_m$ is a scalar tuning parameter, $\Sigma_m$ is the proposal covariance matrix, $\psi^*=(\phi_1^*, \exp(\phi_2^*), \exp(\phi_3^*), \phi_4^*)^T,$ 
$$\pi=\min\left\{1, \frac{\pi(\psi^* \mid {\cal Y}_0^{(0)}, {\cal X}_0^{(0)})\pi_P(\bm{\psi}^{(m-1)})}{\pi(\psi^{(m-1)}\mid {\cal Y}_0^{(0)}, {\cal X}_0^{(0)})\pi_P(\psi^*)}\right\},$$
and $\pi_P(\cdot)$ is the density function of $\psi^*.$ During burn-in, $\Sigma_m$ is updated adaptively using the empirical covariance of recent draws, with a small ridge term added for numerical stability, and $s_m$ is adjusted to maintain a reasonable acceptance rate. 
In our implementation, the MCMC is run for 5{,}000 iterations with a burn-in of 1{,}000 iterations, and posterior inference is based on the retained draws after burn-in. A representative trace plot is provided in the Appendix Figure \ref{fig:traceplot} to illustrate mixing and convergence of the sampler.

For each retained draw $(\mu^{(m)},\tau^{(m)},\nu^{(m)},\alpha^{(m)})$ and a given covariate $\bm{x}_0$, we draw
\[\tilde{\beta}^{(m)}\sim \text{Cauchy}(\mu^{(m)},\tau^{(m)})\]
and compute
\[
\tilde{S}^{(m)}(t\mid \bm{x}_0)
= \exp\left[-t^{\nu^{(m)}}\exp\left\{\alpha^{(m)}+\tilde{\beta}^{(m)}f(\hat{\bm{\theta}}_1, \bm{x}_0)\right\}\right],
\]
 Algorithm \ref{alg:recast_survival} summarizes procedures to sampling survival function for control patients with covariates ${\cal X}_0^{(1)},$ the covaraites of treated patients.

\begin{remark}
To simplify the computation of marginal likelihood (\ref{eq:marginal-lik}), we deploy the transformation
\[
\beta_i=\mu+\tau\tan(u), \qquad u\in(-\pi/2,\pi/2),
\]
which maps a uniform measure on $(-\pi/2,\pi/2)$ into a Gauchy distribution. Thus,
the marginal likelihood function can be written as
\[
\bar{L}(\bm{\psi}\mid\hat{\bm{\theta}}_1, t, \delta, \bm{x} )
=
\frac{1}{\pi}
\int_{-\pi/2}^{\pi/2}
L( \mu+\tau\tan(u), \bm{\psi} \mid \hat{\bm{\theta}}_1, t, \delta, \bm{x})
\, du
\]
and we may approximate this one-dimensional integral using Gauss--Legendre quadrature. In our implementation, we use 64-point Gauss--Legendre quadrature for this numerical integration. 
\end{remark}

\begin{algorithm}[h]
\SetAlgoLined
\DontPrintSemicolon
\caption{Posterior predictive sampling under RECaST-Surv}
\label{alg:recast_survival}
\KwIn{Source estimate $\hat{\bm{\theta}}_1$, time grid $\mathcal{T}$, posterior samples from $\pi(\mu,\tau,\nu,\alpha \mid Data)$, covariate matrix ${\cal X}_0^{(1)}=\{\bm{x}_1^{(1)},\ldots,\bm{x}_{n_{0,1}}^{(1)}\}\}$, number of posterior draws $M$.}
\KwOut{Posterior predictive survival probabilities over $\mathcal{T}$ for the $n_{0,1}$ treated patients.}

\For{$m \leftarrow 1$ \KwTo $M$}{
  Draw $(\mu^{(m)},\tau^{(m)},\nu^{(m)},\alpha^{(m)})$ from the posterior output\;
    \For{$i \leftarrow 1$ \KwTo $n_{0,1}$}{
      Draw $\tilde{\beta}_i^{(m)} \sim \text{Cauchy}(\delta^{(m)},\gamma^{(m)})$\;
      Set the patient-specific linear predictor as $\eta_i^{(m)}=\alpha^{(m)}+\tilde{\beta}_i^{(m)}f(\hat{\bm{\theta}}_1, \bm{x}_i^{(1)})$\;
      \For{$t \in \mathcal{T}$}{
        Compute
        \[
        \tilde{S}_i^{(m)}(t)=\exp\left\{-t^{\nu^{(m)}}\exp(\eta_i^{(m)})\right\}
        \]
      }
    }
    Store the predicted survival curves $\{\tilde{S}_i^{(m)}(t): i=1,\ldots,n_{0,1},\ t\in\mathcal{T}\}$\;
}
\end{algorithm}

\subsection{Treatment Effect Estimation, Hypothesis Testing, and Calibration}

After fitting RECaST-Surv using the current-trial controls and the external source, we apply Algorithm \ref{alg:recast_survival} to the treated patients' covariates to obtain posterior predictive survival curves under the counterfactual control regimen. These predictions are conditional on the observed target data and the source-model estimate, so they may be viewed as patient-specific counterfactual survival curves for the treated patients.

For treated patient $i$ with covariates $\bm{x}_i^{(1)}\in {\cal X}_0^{(1)}$, let $\tilde{S}_{i}^{(m)}(t)$ denote the predicted survival curve from posterior draw $m$. We summarize these posterior predictions by the pointwise median
\[
\tilde{S}_{i}(t)=\text{median}\left\{\tilde{S}_{i}^{(m)}(t): m=1,\ldots,M\right\},
\]
and define the corresponding cumulative hazard function as
\[
\tilde{\Lambda}_{i}(t)=-\log \tilde{S}_{i}(t).
\]
We use the pointwise median rather than the mean because the calibration effect $\beta_i$ follows a Cauchy distribution, and the posterior predictive survival curves may therefore occasionally include extreme draws. The median provides a more stable summary of the counterfactual survival curves and prevents a small number of extreme predictions from dominating the expected event count $E$ and the resulting test statistic defined below.

To compare the observed treated outcomes with these counterfactual control predictions, we use a one-sample log-rank-type procedure based on ${\cal Y}_0^{(1)}$ and $\left\{ \tilde{S}_1(\cdot), \cdots, \tilde{S}_{n_{0,1}}(\cdot)\right\}$. Specifically, let
\[
O=\sum_{i=1}^{n_{0,1}} \delta_{0,i}^{(1)}
\]
be the observed number of events in the treated arm, and define the corresponding expected number of events under the predicted control survival curves as
\[
E=\sum_{i=1}^{n_{0,1}} \tilde{\Lambda}_{i}\left(t_{0,i}^{(1)}\right).
\]
Then, the null hypothesis $H_0$ that the hazard ratio is 1 can be tested based on the test statistic 
\begin{equation}
Z_{lr}=\frac{O-E}{\sqrt{E}}.
\label{eq:test-stat} \end{equation}
The naive p-value can be calculated as $P\left(|N(0, 1)|>|Z_{lr}|\right)$ and the rejection threshold for commonly used two-sided significance level of 0.05 is 1.96.  

In RECaST-Surv, we propose to calibrate this rejection threshold. 
Specifically, we randomly split the control arm into $r+1$ approximately equal folds. In each split, one fold is treated as the training set, while the remaining $r$ folds are combined and treated as a pseudo-treated set. We fit RECaST-Surv using the training fold together with the external source, generate posterior predictive survival function for the pseudo-treated controls, and then compute the naive p-value from the one-sample log-rank-type procedure above.
Suppose this splitting procedure is repeated $B$ times, yielding p-values $p^{(1)},\ldots,p^{(B)}$. We define the calibrated decision threshold at level $\alpha_0$ as
\[
c_{\alpha_0}=\text{the empirical }\alpha_0\text{-quantile of } \{p^{(1)},\ldots,p^{(B)}\}.
\]
In our implementation, we set $B = 100$. To avoid a liberal rejection rule due to finite-sample variability in the calibration step, we further impose a safety cap on the calibrated threshold. Specifically, after obtaining $c_{\alpha_0}$, we use
\[
c_{\alpha_0}^{\mathrm{safe}} = \min(c_{\alpha_0},\,\alpha_0).
\]
 With this calibration step, the null hypothesis is rejected if the naive p-value based on statistic (\ref{eq:test-stat}) is less than $c_{\alpha_0}^{\mathrm{safe}}$.

\begin{remark}
The naive p-value may not be reliable as it ignors the variability of estimated counterfactual survival functions $\tilde{S}_i(\cdot), i=1, \cdots, n_{0,1}.$  The calibration procedure essentially uses a permutation-like procedure to calibrate the null distribution of test statistic. It yeilds a study-specific, data-driven rejection rule based only on the observed control data and matched to the randomization structure of the current trial. Thus, it is a essential step. On the other hand, this calibration does not guarantee exact nominal type I error in finite samples as it can not exactly reproduce the sample sizes used for estimating the counterfactual survival functions and the final test. 
\end{remark}
In addition to testing the between group difference, one may also estimate hazard ratio by 
\[
\widehat{HR}=\frac{O}{E}
\]
under the proportional hazards assumption. The $100(1-\alpha)\%$ confidence interval (CI) for the hazard ratio can be constructed based on the pivotal statistic: 
$$ Z_{HR}=\sqrt{O}\left(\log(O)-\log(E)-\log{HR}\right).$$
In the one-sample setting, $Z_{HR}$ follows a standard Gaussian when the sample size is large, and a $100\times(1-\alpha/2)$\% CI for hazard ratio can be constrcuted as
$$CI_{HR}=\left[\widehat{HR}\exp\left(-\frac{z_{1-\alpha/2}}{\sqrt{O}}\right), \widehat{HR}\exp\left(\frac{z_{1-\alpha/2}}{\sqrt{O}}\right)\right],$$
where $z_{1-\alpha/2}$ is the $(1-\alpha/2)$th quantile of standard normal. Similar to the log-rank test, the distribution of this pivotal statistic may not be exactly standard Gaussian, and we propose to approximate it by the empirical distribution of 
$$Z_{HR}^{(b)}=\left \{\sqrt{O}^{(b)}\left(\log(O^{(b)})-\log(E^{(b)})\right), b=1, \cdots, B\right\}$$
where $(O^{(b)}, E^{(b)})$ are from the $b$th split in the aforementioned claibration procedure, in which the true hazard ratio is 1.  The calibrated CI can then be constructed by replacing $z_{1-\alpha/2}$ by 
$$c_{1-\alpha/2}=\mbox{ the } 100\times(1-\alpha/2) \mbox{th percentile of}~\left \{Z_{HR}^{(1)}, \cdots, Z_{HR}^{(B)}\right\}.$$

In summary, the RECaST-Surv follows the following steps:
\begin{itemize}
\item estimate the structure component based on external controls: $ {\cal Y}_1\cup{\cal X}_1 \to \hat{\bm{\theta}}_1 $
\item estimate the posterior distribution on model parameters based on control group of the target study: $ \hat{\bm{\theta}}_1 ~\mbox{and}~ {\cal Y}_0^{(0)}\cup{\cal X}_0^{(0)} \to \pi\left(\bm{\psi}\mid {\cal Y}_0^{(0)}, {\cal X}_0^{(0)}\right)$
\item estimate the conterfactual survival functions for patients in the treatment group of the target study: $\pi\left(\bm{\psi}\mid {\cal Y}_0^{(0)}, {\cal X}_0^{(0)}\right)~\mbox{ and }~ {\cal X}_0^{(1)} \to \left\{ \tilde{S}_1(\cdot), \cdots, \tilde{S}_{n_{0,1}}(\cdot)\right\}$
\item estimate and test the hazard ratio in the target study: $\left\{ \tilde{S}_1(\cdot), \cdots, \tilde{S}_{n_{0,1}}(\cdot)\right\}~\mbox{and}~{\cal Y}_0^{(1)} \to (\widehat{HR},  Z_{lr}).$
\item calibration: generate CI of HR and p-value for tesitng $HR=1.$
\end{itemize}

\subsection{Extension to Multiple External Sources}

We now extend the design to the case $J>1$. Suppose the same set of covariates is observed across the $J$ external sources, and let $\hat{\bm{\theta}}_j$ denote the source-model estimate obtained from external source $j$, for $j=1,\ldots,J$. We fit RECaST-Surv separately to each pair consisting of external source $j$ and the current-trial control arm, which yields $J$ source-specific fitted models.

For each fitted model, let $\hat{\tau}_j$ denote the posterior median of the scale parameter $\tau$, and let $\tilde{S}_{j,i'}(t)$ denote the resulting posterior predicted survival curve for treated patient $i'$ of source $j$. The use of $\tau$ for weighting is motivated by Lemma 1 of \cite{hickey2024transfer}, which shows, up to first-order approximation, that
\[
\tau \approx \frac{1}{\|\bm{\theta}_s\|^2}
\sqrt{\|\bm{\theta}_s\|^2\|\bm{\theta}_t\|^2-(\bm{\theta}_t^T\bm{\theta}_s)^2}.
\]
A smaller value of $\tau$ indicates that the source parameter vector is more aligned with the target parameter vector, and therefore suggests that the source is a better proxy for the target. This motivates an inverse-$\tau$ weighting scheme.

We define the source weights as
\[
w_j=\frac{1/\hat{\tau}_j}{\sum_{l=1}^J 1/\hat{\tau}_l}, \qquad j=1,\ldots,J,
\]
and combine the source-specific predictions through
\[
\tilde{S}_{i'}(t)=\sum_{j=1}^J w_j \tilde{S}_{j,i'}(t).
\]
These combined survival curves are then used in the same way as in the single-source case to compute the working hazard-ratio estimate, the test statistic, and the calibrated p-value decision rule.

\section{Simulation Studies} \label{sec:sim}

\subsection{Simulation Setup}

We conducted simulation studies to evaluate the performance of the proposed RECaST-Surv design under unequal randomization and varying degrees of compatibility between the trial and external data. In particular, we considered settings where the external data followed the same outcome model as the current trial, as well as settings where the external data contained unmeasured confounding.

\paragraph{Current randomized trial ($j=0$).}
We generated the current trial with total sample size $n_0 \in \{200,300\}$. Patients were randomized to treatment or control under an $r:1$ allocation, where $r \in \{1,2,3\}$.

For each patient $i$, we generated $d = 3$ baseline covariates
\[
\bm{x}_{0,i} = (x_{0,i,1},x_{0,i,2},x_{0,i,3}),
\]
with
\[
x_{0,i,d} \sim N(0,1), \qquad d=1,2,3.
\]
Survival times were generated from a Weibull proportional hazards model with shape parameter $\nu=1.2$ and baseline scale parameter $\lambda=0.01$. Specifically, recall that we used the super-script $k = 0, 1$ for the control and treated arms, respectively. The hazard function was
\[
\lambda^{(0)}(t \mid \bm{x}_{0,i})
=
\nu t^{\nu-1}\exp(\alpha+\bm{x}_{0,i}^T\bm{\theta}_0 )
~
\mbox{ and }
~
\lambda^{(1)}(t \mid \bm{x}_{0,i})
=
\nu t^{\nu-1}\exp(\alpha+\bm{x}_{0,i}^T\bm{\theta}_0+\Delta)
\]
for control patient and treated patient, respectively, 
where
\[
\bm{\theta}_0 = (0.2,0.3,0.4).
\]
We considered three treatment-effect settings, corresponding to
\[
\Delta \in \{0,-0.25,-0.5\}.
\]
These yield true hazard ratios
\[
HR = \exp(\Delta) \in \{1, 0.779, 0.607\}.
\]
Thus, the case $\Delta=0$ corresponds to the null setting, while $\Delta=-0.25$ and $\Delta=-0.5$ correspond to two alternative settings with increasing treatment benefit. Administrative censoring was imposed at $T=60$ months, and additional random dropout was calibrated to give approximately 15\% censoring overall.

\paragraph{External data sources ($j=1,2,3$).}
We generated three external data sources with sample sizes
\[
n_1=100,\qquad n_2=250,\qquad n_3=400.
\]
All external subjects were controls.
As in the current trial, the observed external covariates were generated from $N(0,1)$.

We considered two scenarios for the external data-generating mechanism.
\begin{itemize}
    \item \textbf{Scenario A: congruent models (``Same'').}  
    In this setting, the external data followed the same structural model as the current trial, in the sense that outcomes depended on the same three observed covariates. To allow for different degrees of compatibility across sources, we used different coefficient vectors:
    \begin{itemize}
        \item \textbf{Source 1:} $\bm{\theta}_1 = (0.2,0.3,0.4)$,
        \item \textbf{Source 2:} $\bm{\theta}_2 = (0.15,0.35,0.45)$,
        \item \textbf{Source 3:} $\bm{\theta}_3 = (0.22,0.28,0.38)$.
    \end{itemize}
    Source 1 is identical to the trial model, while Sources 2 and 3 are close but not identical.

    \item \textbf{Scenario B: incongruent models (``Diff'').}  
    In this setting, we introduced an unmeasured confounder to the external data. Specifically, for each external subject we generated
    \[
    x_{j,i,4} \sim N(0,1), \qquad j=1,2,3,
    \]
    and generated outcomes from a Weibull model depending on four covariates, while the analysis model only used the three observed covariates $(x_{j,i,1},x_{j,i,2},x_{j,i,3})$. The coefficient vectors in the external sources were
    \begin{itemize}
        \item \textbf{Source 1:} $\bm{\theta}_1 = (0.2,0.3,0.4,0.5)$,
        \item \textbf{Source 2:} $\bm{\theta}_2 = (0.15,0.35,0.45,0.55)$,
        \item \textbf{Source 3:} $\bm{\theta}_3 = (0.22,0.28,0.38,0.51)$.
    \end{itemize}
    This scenario was used to assess robustness when the external data are not fully compatible with the current trial and contain unmeasured confounding.
\end{itemize}

For each combination of $(n_0,r,\Delta)$ and external-data scenario, we generated 1{,}000 simulated datasets. We compared the proposed RECaST-Surv method with two alternatives: the RCT-only analysis without borrowing (denoted by RCT) and the robust meta-analytic-predictive prior (R-MAP). We report the estimated hazard ratio, its standard deviation (SD), mean squared error (MSE), and the rejection rate, which corresponds to type I error under the null and power under the alternatives. 

For R-MAP, we also report a calibrated version, denoted by C-RMAP. This uses the same point estimates as R-MAP, so the estimated hazard ratio, SD, and MSE are unchanged. The difference is only in the rejection threshold. Specifically, for each scenario, we used the simulated null datasets (HR $=1$) to choose a cutoff that yields an empirical type I error rate of 5\%. We then applied this calibrated cutoff to the alternative settings (HR $=0.779$ and HR $=0.607$) to evaluate the corresponding power.

To characterize the amount of event information available in the current trial, Appendix~\ref{tab:expected_events} reports the expected numbers of treated and control events for each trial sample size, randomization ratio, and treatment-effect setting. As expected, increasing $r$ increases the expected number of treated events but reduces the number of concurrent control events, which is the setting in which external control information is most relevant.

\subsection{Simulation Results}

Tables \ref{tab:simulation_results_same} and \ref{tab:simulation_results_diff} summarize the simulation results under the congruent (``Same'') and incongruent (``Diff'') scenarios, respectively.

\begin{table}[p]
\centering
\caption{Simulation results under the congruent (``Same'') scenario. For each true hazard ratio, $HR \in \{1.000, 0.779, 0.607\}$, we report the estimated hazard ratio (Est.), standard deviation (SD), mean squared error (MSE), and rejection rate, where the rejection rate corresponds to the type I error rate when $HR=1$ and to power otherwise. ``C-RMAP'' denotes the calibrated version of R-MAP, obtained by choosing the rejection threshold so that the empirical type I error rate is 5\% under the null. Because the calibration only changes the decision threshold, the point estimates, SDs, and MSEs are unchanged from R-MAP.}
\label{tab:simulation_results_same}
\begingroup\small
\setlength{\tabcolsep}{2.5pt}
\renewcommand{\arraystretch}{0.9}
\begin{tabular}{@{}lcl|cccc|cccc|cccc@{}}
\toprule
 &  &  & \multicolumn{4}{c}{HR $=1.000$} & \multicolumn{4}{c}{HR $=0.779$} & \multicolumn{4}{c}{HR $=0.607$} \\
\cmidrule(lr){4-7}\cmidrule(lr){8-11}\cmidrule(lr){12-15}
\textbf{N} & \textbf{r} & \textbf{Method} & \textbf{Est.} & \textbf{SD} & \textbf{MSE} & \textbf{TIER} & \textbf{Est.} & \textbf{SD} & \textbf{MSE} & \textbf{Power} & \textbf{Est.} & \textbf{SD} & \textbf{MSE} & \textbf{Power} \\
\midrule
\multirow{12}{*}{200} & \multirow{4}{*}{1} & RCT & 1.019 & 0.178 & 0.032 & 0.047 & 0.818 & 0.147 & 0.023 & 0.232 & 0.653 & 0.121 & 0.017 & 0.665 \\
 &  & RECaST-Surv & 1.029 & 0.166 & 0.028 & 0.035 & 0.826 & 0.138 & 0.021 & 0.365 & 0.659 & 0.115 & 0.016 & 0.837 \\
 &  & RMAP & 0.968 & 0.231 & 0.054 & 0.080 & 0.785 & 0.176 & 0.031 & 0.414 & 0.625 & 0.155 & 0.024 & 0.899 \\
 &  & C-RMAP & -- & -- & -- & 0.050 & -- & -- & -- & 0.127 & -- & -- & -- & 0.560 \\
\cmidrule{2-15}
 & \multirow{4}{*}{2} & RCT & 1.030 & 0.195 & 0.039 & 0.048 & 0.811 & 0.150 & 0.024 & 0.225 & 0.649 & 0.123 & 0.017 & 0.657 \\
 &  & RECaST-Surv & 1.038 & 0.186 & 0.036 & 0.046 & 0.822 & 0.140 & 0.021 & 0.337 & 0.658 & 0.115 & 0.016 & 0.804 \\
 &  & RMAP & 0.982 & 0.232 & 0.054 & 0.072 & 0.793 & 0.184 & 0.034 & 0.419 & 0.634 & 0.155 & 0.025 & 0.874 \\
 &  & C-RMAP & -- & -- & -- & 0.050 & -- & -- & -- & 0.237 & -- & -- & -- & 0.762 \\
\cmidrule{2-15}
 & \multirow{4}{*}{3} & RCT & 1.019 & 0.201 & 0.041 & 0.043 & 0.824 & 0.175 & 0.033 & 0.199 & 0.658 & 0.142 & 0.023 & 0.552 \\
 &  & RECaST-Surv & 1.042 & 0.194 & 0.039 & 0.036 & 0.836 & 0.166 & 0.031 & 0.256 & 0.668 & 0.136 & 0.022 & 0.680 \\
 &  & RMAP & 0.984 & 0.240 & 0.058 & 0.070 & 0.789 & 0.203 & 0.041 & 0.388 & 0.639 & 0.152 & 0.024 & 0.864 \\
 &  & C-RMAP & -- & -- & -- & 0.050 & -- & -- & -- & 0.177 & -- & -- & -- & 0.676 \\
\midrule
\multirow{12}{*}{300} & \multirow{4}{*}{1} & RCT & 1.012 & 0.140 & 0.020 & 0.035 & 0.809 & 0.116 & 0.014 & 0.324 & 0.647 & 0.096 & 0.011 & 0.847 \\
 &  & RECaST-Surv & 1.019 & 0.130 & 0.017 & 0.052 & 0.814 & 0.108 & 0.013 & 0.512 & 0.651 & 0.091 & 0.010 & 0.937 \\
 &  & RMAP & 0.983 & 0.201 & 0.041 & 0.070 & 0.785 & 0.178 & 0.032 & 0.522 & 0.633 & 0.140 & 0.020 & 0.949 \\
 &  & C-RMAP & -- & -- & -- & 0.050 & -- & -- & -- & 0.416 & -- & -- & -- & 0.909 \\
\cmidrule{2-15}
 & \multirow{4}{*}{2} & RCT & 1.016 & 0.156 & 0.025 & 0.055 & 0.807 & 0.125 & 0.016 & 0.313 & 0.648 & 0.103 & 0.012 & 0.809 \\
 &  & RECaST-Surv & 1.026 & 0.146 & 0.022 & 0.046 & 0.815 & 0.115 & 0.015 & 0.465 & 0.655 & 0.098 & 0.012 & 0.910 \\
 &  & RMAP & 0.977 & 0.211 & 0.045 & 0.075 & 0.786 & 0.177 & 0.031 & 0.516 & 0.631 & 0.143 & 0.021 & 0.951 \\
 &  & C-RMAP & -- & -- & -- & 0.050 & -- & -- & -- & 0.393 & -- & -- & -- & 0.908 \\
\cmidrule{2-15}
 & \multirow{4}{*}{3} & RCT & 1.012 & 0.167 & 0.028 & 0.046 & 0.813 & 0.140 & 0.021 & 0.289 & 0.647 & 0.111 & 0.014 & 0.745 \\
 &  & RECaST-Surv & 1.027 & 0.155 & 0.025 & 0.051 & 0.824 & 0.131 & 0.019 & 0.397 & 0.656 & 0.103 & 0.013 & 0.879 \\
 &  & RMAP & 0.973 & 0.217 & 0.048 & 0.066 & 0.784 & 0.179 & 0.032 & 0.525 & 0.631 & 0.144 & 0.021 & 0.936 \\
 &  & C-RMAP & -- & -- & -- & 0.050 & -- & -- & -- & 0.337 & -- & -- & -- & 0.853 \\
\bottomrule
\end{tabular}%
\endgroup
\end{table}

\begin{table}[p]
\centering
\caption{Simulation results under the incongruent (``Diff'') scenario. For each true hazard ratio, $HR \in \{1.000, 0.779, 0.607\}$, we report the estimated hazard ratio (Est.), standard deviation (SD), mean squared error (MSE), and rejection rate, where the rejection rate corresponds to the type I error rate when $HR=1$ and to power otherwise. ``C-RMAP'' denotes the calibrated version of R-MAP, obtained by choosing the rejection threshold so that the empirical type I error rate is 5\% under the null. Because the calibration only changes the decision threshold, the point estimates, SDs, and MSEs are unchanged from R-MAP.}
\label{tab:simulation_results_diff}
\begingroup\small
\setlength{\tabcolsep}{3pt}
\renewcommand{\arraystretch}{0.9}
\begin{tabular}{@{}lcl|cccc|cccc|cccc@{}}
\toprule
 &  &  & \multicolumn{4}{c}{HR $=1.000$} & \multicolumn{4}{c}{HR $=0.779$} & \multicolumn{4}{c}{HR $=0.607$} \\
\cmidrule(lr){4-7}\cmidrule(lr){8-11}\cmidrule(lr){12-15}
\textbf{N} & \textbf{r} & \textbf{Method} & \textbf{Est.} & \textbf{SD} & \textbf{MSE} & \textbf{TIER} & \textbf{Est.} & \textbf{SD} & \textbf{MSE} & \textbf{Power} & \textbf{Est.} & \textbf{SD} & \textbf{MSE} & \textbf{Power} \\
\midrule
\multirow{12}{*}{200} & \multirow{4}{*}{1} & RCT & 1.010 & 0.172 & 0.030 & 0.051 & 0.810 & 0.143 & 0.021 & 0.225 & 0.647 & 0.118 & 0.016 & 0.697 \\
 &  & RECaST-Surv & 1.022 & 0.164 & 0.028 & 0.041 & 0.819 & 0.136 & 0.020 & 0.390 & 0.654 & 0.113 & 0.015 & 0.838 \\
 &  & RMAP & 0.979 & 0.234 & 0.055 & 0.070 & 0.791 & 0.189 & 0.036 & 0.400 & 0.635 & 0.159 & 0.026 & 0.873 \\
 &  & C-RMAP & -- & -- & -- & 0.050 & -- & -- & -- & 0.196 & -- & -- & -- & 0.689 \\
\cmidrule{2-15}
 & \multirow{4}{*}{2} & RCT & 1.020 & 0.188 & 0.036 & 0.047 & 0.816 & 0.153 & 0.025 & 0.226 & 0.651 & 0.124 & 0.017 & 0.620 \\
 &  & RECaST-Surv & 1.031 & 0.174 & 0.031 & 0.050 & 0.825 & 0.142 & 0.022 & 0.329 & 0.658 & 0.115 & 0.016 & 0.798 \\
 &  & RMAP & 0.975 & 0.247 & 0.062 & 0.084 & 0.792 & 0.181 & 0.033 & 0.406 & 0.635 & 0.155 & 0.025 & 0.873 \\
 &  & C-RMAP & -- & -- & -- & 0.050 & -- & -- & -- & 0.129 & -- & -- & -- & 0.469 \\
\cmidrule{2-15}
 & \multirow{4}{*}{3} & RCT & 1.021 & 0.204 & 0.042 & 0.050 & 0.817 & 0.166 & 0.029 & 0.201 & 0.651 & 0.135 & 0.020 & 0.582 \\
 &  & RECaST-Surv & 1.042 & 0.199 & 0.041 & 0.039 & 0.833 & 0.161 & 0.029 & 0.266 & 0.665 & 0.131 & 0.021 & 0.700 \\
 &  & RMAP & 0.977 & 0.248 & 0.062 & 0.087 & 0.781 & 0.211 & 0.045 & 0.401 & 0.635 & 0.159 & 0.026 & 0.849 \\
 &  & C-RMAP & -- & -- & -- & 0.050 & -- & -- & -- & 0.091 & -- & -- & -- & 0.388 \\
\midrule
\multirow{12}{*}{300} & \multirow{4}{*}{1} & RCT & 1.012 & 0.146 & 0.021 & 0.056 & 0.809 & 0.121 & 0.015 & 0.339 & 0.645 & 0.099 & 0.011 & 0.847 \\
 &  & RECaST-Surv & 1.018 & 0.137 & 0.019 & 0.055 & 0.814 & 0.113 & 0.014 & 0.489 & 0.650 & 0.093 & 0.010 & 0.944 \\
 &  & RMAP & 0.966 & 0.230 & 0.054 & 0.083 & 0.785 & 0.175 & 0.031 & 0.541 & 0.633 & 0.136 & 0.019 & 0.955 \\
 &  & C-RMAP & -- & -- & -- & 0.050 & -- & -- & -- & 0.184 & -- & -- & -- & 0.716 \\
\cmidrule{2-15}
 & \multirow{4}{*}{2} & RCT & 1.015 & 0.150 & 0.023 & 0.043 & 0.814 & 0.123 & 0.016 & 0.287 & 0.650 & 0.100 & 0.012 & 0.807 \\
 &  & RECaST-Surv & 1.024 & 0.142 & 0.021 & 0.048 & 0.821 & 0.116 & 0.015 & 0.451 & 0.656 & 0.096 & 0.012 & 0.920 \\
 &  & RMAP & 0.982 & 0.206 & 0.043 & 0.071 & 0.793 & 0.164 & 0.027 & 0.518 & 0.635 & 0.140 & 0.020 & 0.948 \\
 &  & C-RMAP & -- & -- & -- & 0.050 & -- & -- & -- & 0.384 & -- & -- & -- & 0.886 \\
\cmidrule{2-15}
 & \multirow{4}{*}{3} & RCT & 1.007 & 0.166 & 0.027 & 0.053 & 0.807 & 0.137 & 0.020 & 0.289 & 0.646 & 0.112 & 0.014 & 0.778 \\
 &  & RECaST-Surv & 1.019 & 0.159 & 0.026 & 0.055 & 0.816 & 0.132 & 0.019 & 0.420 & 0.653 & 0.108 & 0.014 & 0.860 \\
 &  & RMAP & 0.982 & 0.214 & 0.046 & 0.061 & 0.788 & 0.179 & 0.032 & 0.501 & 0.642 & 0.130 & 0.018 & 0.923 \\
 &  & C-RMAP & -- & -- & -- & 0.050 & -- & -- & -- & 0.359 & -- & -- & -- & 0.852 \\
\bottomrule
\end{tabular}%
\endgroup
\end{table}

Across all settings, the estimated hazard ratios are reasonably close to the nominal treatment effects used in the data-generating model for all methods. Small departures from the target values $HR=\exp(\Delta)$ are expected, because the Weibull data-generating mechanism also includes prognostic covariates and censoring; thus, non-collapsibility and dilution play roles here, and the empirical hazard-ratio estimates in the table are not expected to be centered exactly at the conditional treatment effect, even for the RCT-only analysis. Relative to the RCT-only analysis, RECaST-Surv tends to produce slightly larger hazard-ratio estimates, while R-MAP tends to produce slightly smaller ones. These differences are small, and there is no indication that RECaST-Surv design introduces substantial bias.

In terms of estimation accuracy, RECaST-Surv design generally has smaller SD and MSE than the RCT-only analysis, indicating a gain in precision from information borrowing. This improvement is more noticeable when the total trial sample size is larger. In contrast, when the total sample size is fixed, increasing the randomization ratio $r$ generally reduces precision for all methods, since fewer patients remain in the concurrent control arm. This pattern is expected in our setting because the current-trial control arm is the target dataset used to calibrate the external information.

For operating characteristics, RECaST-Surv design yields type I error rates close to the nominal 5\% level in both the ``Same'' and ``Diff'' scenarios, while providing a clear gain in power over the RCT-only analysis. R-MAP often achieves even higher power, but this comes with inflated type I error in a number of settings. After calibrating R-MAP to 5\% type I error through C-RMAP, its power decreases substantially. In comparison, RECaST-Surv design generally has higher power than C-RMAP while maintaining type I error near the nominal level.

The power gain of RECaST-Surv design is seen in both the congruent and incongruent scenarios. This suggests that the proposed method remains stable even when the external data are not fully compatible with the current trial, including the setting where the external sources contain an unmeasured confounder. Overall, the simulation results suggest that RECaST-Surv design provides a favorable balance between precision, power, and type I error control in unequal randomized trials with external information borrowing.

\begin{figure}[p]
  \centering
  \includegraphics[width=0.85\textwidth]{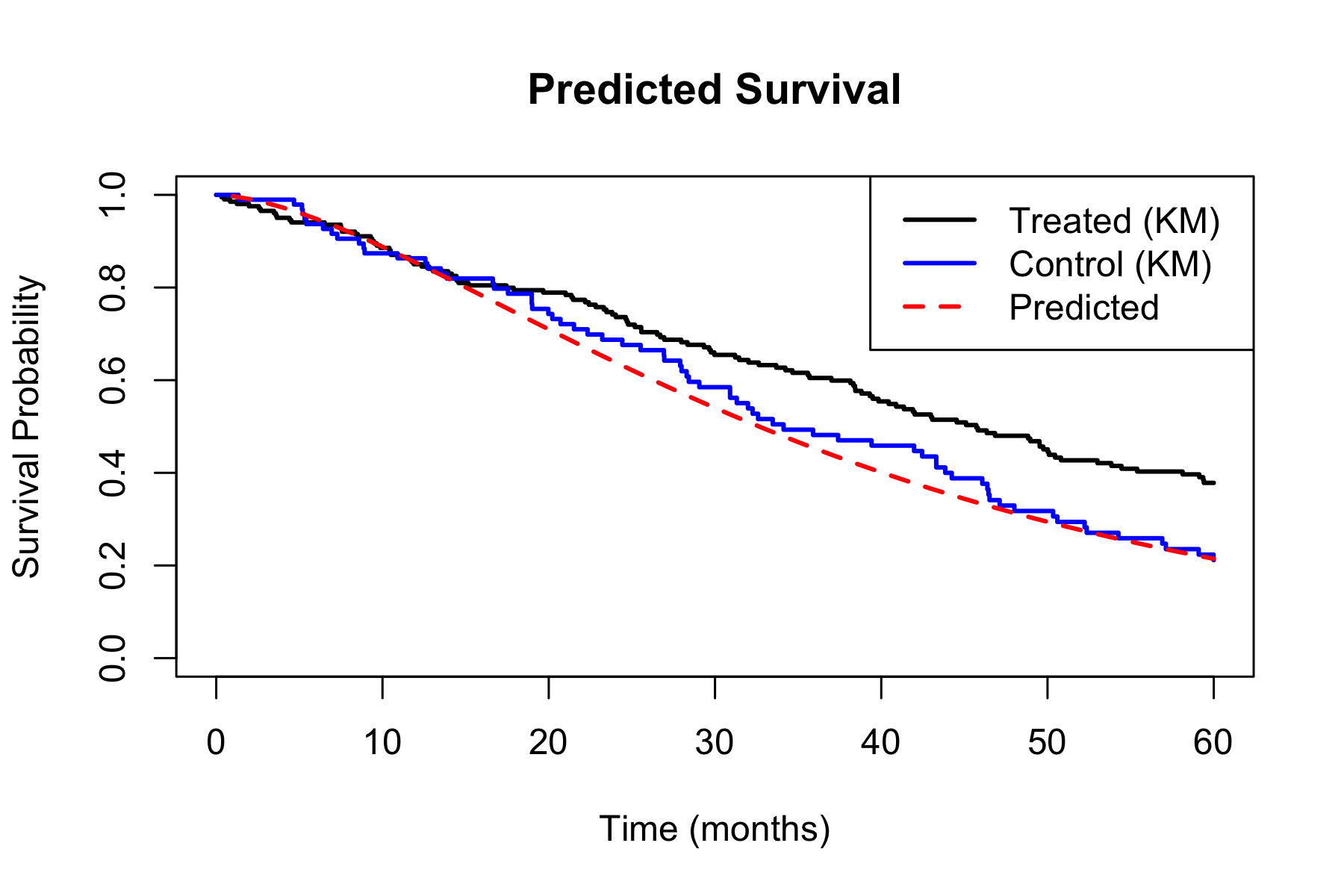}
  \caption{Predicted counterfactual survival curve from one randomly selected simulated dataset under the congruent (``Same'') scenario, with $N=300$, $r=2$, and true $HR=0.607$. The black and blue curves are the Kaplan--Meier curves for the observed treated and control patients, respectively. The red dashed curve is the RECaST-Surv predicted counterfactual control survival curve for the treated patients.}
  \label{fig:predicted_survival_example}
\end{figure}

As a visual illustration of the proposed procedure at the trial level, we randomly selected one simulated dataset from the congruent (``Same'') scenario with $N=300$, $r=2$, and true $HR=0.607$. Figure \ref{fig:predicted_survival_example} shows the observed Kaplan--Meier curves for the treated and control patients, together with the RECaST-Surv predicted counterfactual control survival curve for the treated patients. In this example, the treated-arm curve lies above both the observed control curve and the predicted counterfactual control curve, consistent with a treatment benefit in this simulated trial. The observed number of events in the treated arm was 115, whereas the expected number of events under the predicted counterfactual control survival curve for those treated patients was 176.86. Accordingly, the estimated hazard ratio was $0.650$, with a calibrated 95\% confidence interval of $(0.503, 0.841)$ based on the calibrated critical value $c_{1-\alpha/2}=2.771$. 
The trial-level $p$-value was $1.65\times 10^{-6}$, which was below the calibrated rejection threshold of $0.0056$. Thus, this example leads to rejection of the null hypothesis and illustrates how the predicted counterfactual survival curve is combined with the calibrated testing rule in one simulated trial.

\subsection{Sensitivity Analysis}\label{subsec:sens}

To further assess the robustness of the proposed RECaST-Surv design, we conducted a series of sensitivity analyses under more challenging data-generating settings. These analyses were designed to examine how the method behaves when the main assumptions are stressed.

Specifically, we considered four aspects.

\begin{itemize}
    \item[1).] \textbf{Set A: Sample size imbalance and coefficient magnitude.}  
    We varied the external sample sizes from $50$ to $500$ and considered settings where the external coefficients were systematically larger or smaller than those in the current trial. We also examined mixed settings in which sources with larger or smaller sample sizes had different coefficient magnitudes.

    \item[2).] \textbf{Set B: Covariate correlation and collinearity.}  
    We introduced correlation among the baseline covariates with $\rho \in \{-0.5,0.5\}$ to assess stability under collinearity. Both symmetric settings, where the trial and external data shared the same correlation structure, and asymmetric settings, where they differed, were considered.

    \item[3).] \textbf{Set C: Parameter transmission reliability.}  
    We considered scenarios where the estimated coefficients from the external sources were corrupted during transmission, either through a systematic shift ($\pm 0.5$) or through substantial random noise (standard deviation equal to 1.0). This was meant to reflect administrative errors or unstable source-model estimation.

    \item[4).] \textbf{Set D: Covariate distributional shift.}  
    We used rejection sampling to generate settings with disjoint covariate support between the trial and external sources, for example $X_1>0$ in the current trial and $X_1<0$ in the external data. This represents an extreme form of population heterogeneity.
\end{itemize}

\begin{figure}[p]
  \centering
  \begin{subfigure}[b]{0.48\textwidth}
    \centering
    \includegraphics[width=\textwidth]{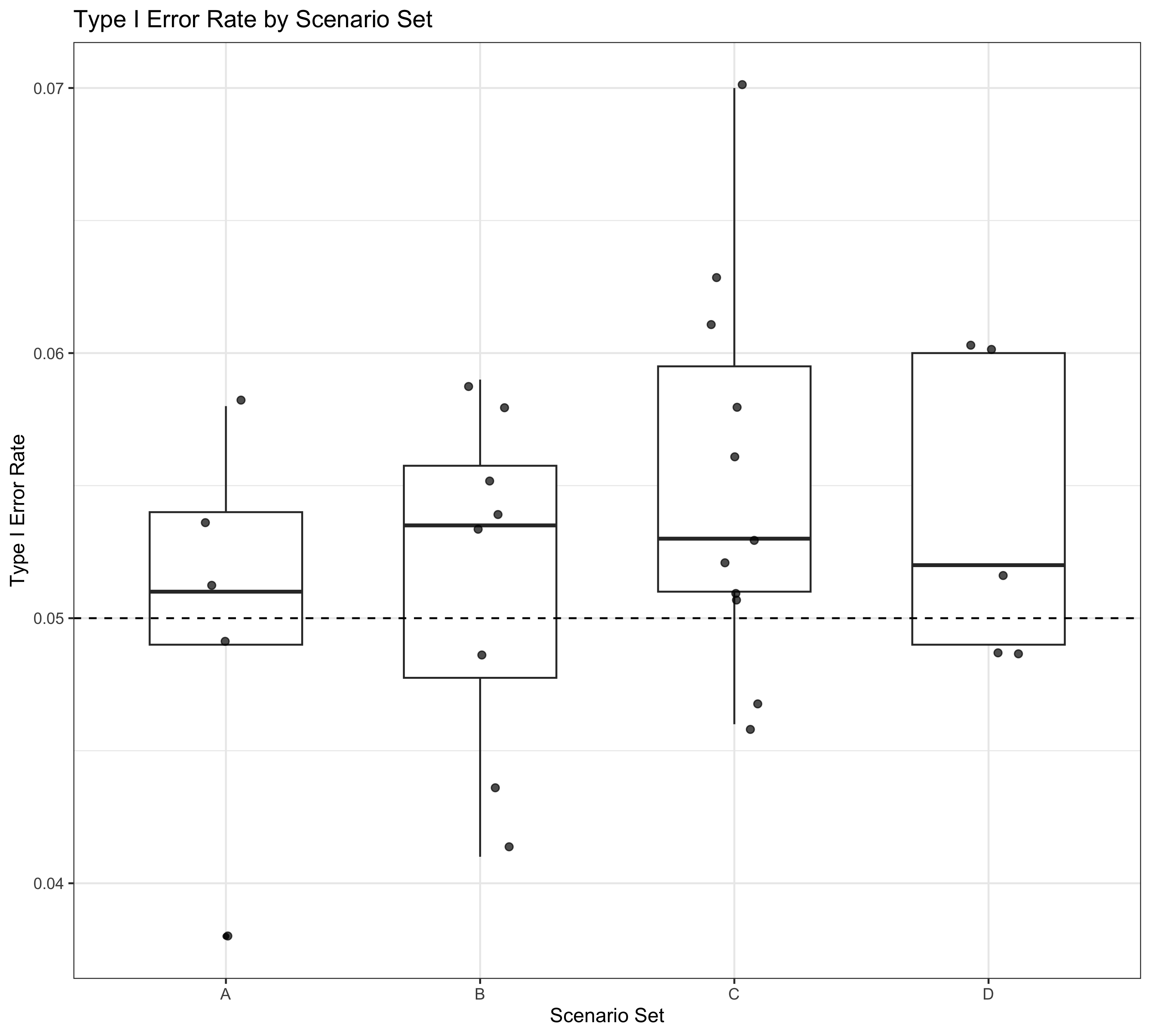}
    \caption{Type I error rate}
    \label{fig:tier_sens}
  \end{subfigure}
  \hfill
  \begin{subfigure}[b]{0.48\textwidth}
    \centering
    \includegraphics[width=\textwidth]{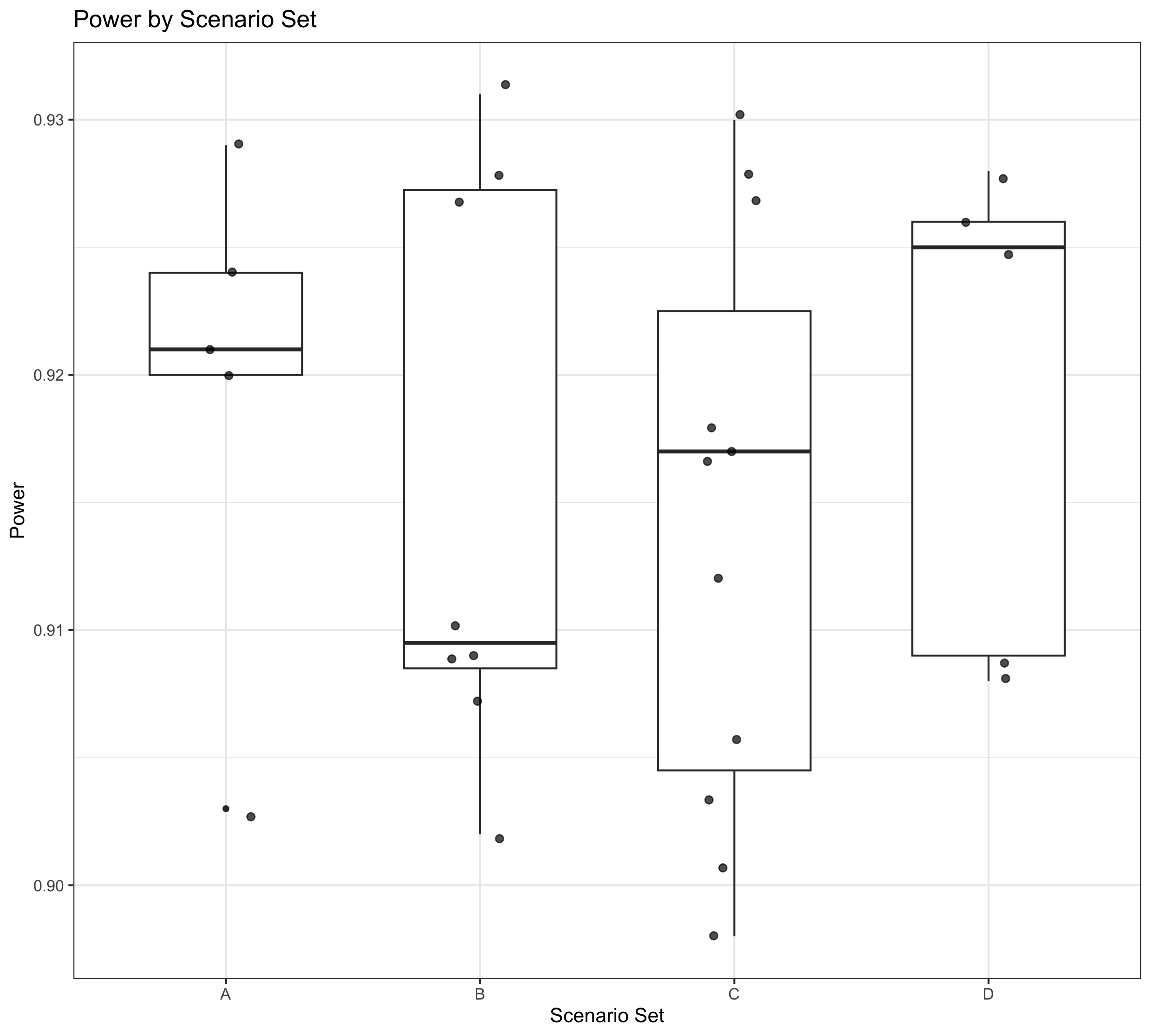}
    \caption{Power}
    \label{fig:power_sens}
  \end{subfigure}
  \caption{Sensitivity analysis results across two simulation sets. Set A considers external sample size and coefficient magnitude, and Set B considers covariate correlation. Set C considers parameter corruption, and Set D considers biased sampling (covariate shift).}
  \label{fig:sens_results}
\end{figure}

Additional details are given in Appendix \ref{subsec:sens_ana}. Across these settings, RECaST-Surv design remained stable. As shown in Figure \ref{fig:sens_results}, panels \subref{fig:tier_sens} and \subref{fig:power_sens}, the type I error rate stayed within a reasonable range, from 0.038 to 0.070, and the method continued to show a power gain of roughly 10\%--12\% over the RCT-only analysis (ranging from 0.898 to 0.931), even in challenging settings where the external coefficients were consistently larger or smaller than the trial truth. These results suggest that the transfer-learning calibration in RECaST-Surv design can adapt to the scale and quality of the external information without becoming overly aggressive.

\section{Real Data Analysis}\label{sec:data_ana}

To examine the performance of RECaST-Surv design in a realistic setting, we applied the proposed method to patient-level data from the Netherlands ALS Registry \citep{van2023hybrid}. Amyotrophic lateral sclerosis (ALS) is a rapidly progressive neurodegenerative disease, so efficient trial designs are of particular interest in this setting.

\subsection{Dataset and Preprocessing}

We used patient-level data from $N=2{,}790$ patients in the Netherlands ALS Registry. The outcome is time-to-event. To mimic the baseline information commonly available in ALS clinical trials, we considered seven baseline covariates: sex, age, body mass index (BMI), forced vital capacity (FVC), disease duration, ALS Functional Rating Scale-Revised (ALSFRS-R) total score, and ALSFRS-R slope. All covariates were centered and scaled before analysis.

\paragraph{Trial emulation and external sources.}
We used a resampling-based procedure to emulate a current trial together with multiple external sources. In each of 1{,}000 iterations, we first sampled 300 patients to represent the current trial. These 300 patients were then randomized in a $2:1$ ratio, with $n_{0,1}=200$ patients assigned to a pseudo-treatment arm and $n_{0,0}=100$ patients assigned to the pseudo-concurrent control arm. The remaining 2{,}490 patients were treated as external data and were randomly divided into three external sources with sample sizes $n_1=500$, $n_2=1{,}000$, and $n_3=990$.

We considered two trial-cohort sampling settings. In the first setting, the 300 trial patients were sampled uniformly from the registry. This setting represents the case where the emulated current trial and the external sources arise from the same underlying registry population. In the second setting, we used a biased sampling mechanism to make the current trial population different from the remaining external population. This was motivated by the practical concern that patients enrolled in a trial may not be a random subset of the registry population. For example, a trial may preferentially enroll younger patients, or may underrepresent certain patients because of eligibility criteria, disease severity, or enrollment feasibility.

Specifically, for each candidate patient $i$, we defined a sampling weight $\omega_i$ for inclusion in the current trial as
\[
\omega_i \propto \exp\{-0.5 \cdot \mathrm{AGE}_i + 0.5 \cdot \mathrm{DISDUR}_i\},
\]
where $\mathrm{AGE}_i$ and $\mathrm{DISDUR}_i$ denote the standardized age and disease duration variables. Because the covariates were standardized before sampling, this weighting scheme upweights younger patients and patients with longer disease duration. To avoid having a few patients dominate the sampling distribution, the weights were truncated at 20 times the median weight. The current trial cohort was then sampled without replacement according to these weights. After the trial cohort was selected, the remaining patients were divided into the three external sources in the same way as in the random sampling setting.

Because all patients in the registry received standard of care, the observed data correspond to a null setting in which there is no treatment effect. We therefore considered two outcome scenarios under each trial-cohort sampling setting.

\paragraph{Null scenario.}
In the null scenario, we retained the observed survival times and censoring indicators for all patients. Since the pseudo-treatment and pseudo-control groups were sampled from the same registry without any intervention, the true hazard ratio should be around 1. This setting was used to evaluate the empirical type I error.

\paragraph{Alternative scenario.}
In the alternative scenario, we preserved the sampled trial cohort and treatment assignment from the corresponding null dataset, and introduced a nonzero treatment effect exclusively in the pseudo-treated arm.Specifically, let $\Delta$ denote the target log hazard ratio, with $\Delta=-0.5$ corresponding to $HR \approx 0.607$. For each pseudo-treated patient, we computed a linear predictor $\bm{x}_i^T\hat{\bm{\beta}}$ using the coefficient estimates $\hat{\bm{\beta}}$ from a global Cox model fitted to the full standardized registry. A new event time was then generated by sampling a uniform random variable $U_i$ and solving
\[
\hat{\Lambda}_0(T_i)\exp(\bm{x}_i^T\hat{\bm{\beta}}+\Delta) = U_i,
\]
where $\hat{\Lambda}_0(\cdot)$ is the estimated baseline cumulative hazard function from the Cox model and $U_i$ follows an unit exponential distribution. Thus, the treated patients' outcomes were replaced by event times generated under the target proportional hazards effect, while preserving the covariate structure learned from the real data.

The pseudo-control arm was left unchanged and retained its original observed outcomes. For the pseudo-treated arm, censoring times were generated independently from an exponential distribution and then combined with administrative censoring at the pre-specified time horizon (in this case, it is set to 10 years). Therefore, the data generated under the alternative differ from those under the null only through the treated-arm outcomes, while preserving the same sampled patients and treatment assignment.



For both scenarios, RECaST-Surv design was compared with the RCT-only analysis that used only the concurrent control arm and did not borrow from the external data. As in the simulation study, we report the estimated hazard ratio, standard deviation, mean squared error (MSE), and the rejection rate, which corresponds to type I error under the null and power under the alternative. For RECaST-Surv design, the posterior predictive survival curves and the resulting test statistic were evaluated over a time horizon of 10 years (120 months).

\subsection{Analysis Results}

The results from the ALS trial emulation are summarized in Table \ref{tab:real_data_results}.


\begin{table}[ht]
\centering
\caption{Results from the ALS trial emulation study comparing RECaST-Surv with the standard RCT analysis without borrowing. The table reports the estimated hazard ratio (Est. HR), standard deviation (SD), mean squared error (MSE), and rejection rate, where the rejection rate is the type I error rate under the null and power under the alternative. Results are based on 1{,}000 resampling iterations.}
\label{tab:real_data_results}
\resizebox{\textwidth}{!}{%
\begin{tabular}{ll | cccc | cccc}
\toprule
 &  & \multicolumn{4}{c}{\textbf{Null Scenario ($H_0: HR = 1.000$)}} 
 & \multicolumn{4}{c}{\textbf{Alt. Scenario ($H_1: HR = 0.607$)}} \\
\cmidrule(lr){3-6} \cmidrule(lr){7-10}
\textbf{Sampling Setting} & \textbf{Method} 
& \textbf{Est.} & \textbf{SD} & \textbf{MSE} & \textbf{TIER} 
& \textbf{Est.} & \textbf{SD} & \textbf{MSE} & \textbf{Power} \\
\midrule
\multirow{2}{*}{Random sampling}
& RECaST-Surv        & 1.061 & 0.138 & 0.023 & 0.063 & 0.686 & 0.103 & 0.016 & 0.957 \\
& RCT (No Borrowing) & 1.008 & 0.128 & 0.016 & 0.055 & 0.684 & 0.094 & 0.015 & 0.828 \\
\midrule
\multirow{2}{*}{Biased sampling}
& RECaST-Surv        & 1.141 & 0.144 & 0.040 & 0.025 & 0.708 & 0.097 & 0.020 & 0.939 \\
& RCT (No Borrowing) & 1.007 & 0.131 & 0.017 & 0.045 & 0.680 & 0.092 & 0.014 & 0.812 \\
\bottomrule
\end{tabular}%
}
\end{table}

Under the random sampling setting, both methods produced hazard ratio estimates close to 1 under the null scenario. The RCT-only analysis yielded a type I error rate of 5.5\%, while RECaST-Surv yielded 6.3\%. Thus, in this real-data example, RECaST-Surv exhibits only mild type I error inflation, consistent with the simulation findings.

Under the alternative scenario with random sampling, the two methods produced nearly identical hazard ratio estimates. The estimated hazard ratio was 0.686 for RECaST-Surv and 0.684 for the RCT-only analysis, with comparable MSE values. These results indicate that the proposed design achieves estimation performance similar to that of the no-borrowing analysis in this setting.

The primary advantage of RECaST-Surv in this application lies in hypothesis testing. Under the alternative scenario with random sampling, the proposed design achieved a power of 95.7\%, compared with 82.8\% for the RCT-only analysis. This substantial gain in power is obtained by borrowing information from the external registry, with only a modest increase in type I error.

The biased sampling setting provides a more challenging emulation, because the current trial cohort is no longer sampled as a random subset of the registry. In this setting, RECaST-Surv produced a slightly larger estimated hazard ratio under the null, 1.141, compared with 1.007 for the RCT-only analysis. However, the empirical type I error remained controlled at 2.5\% for RECaST-Surv and 4.5\% for the RCT-only analysis. This suggests that, although biased trial sampling can introduce some shift in the point estimate when external information is borrowed, the calibrated testing procedure remains conservative in this setting.

Under the alternative scenario with biased sampling, RECaST-Surv again achieved a clear power gain. The power was 93.9\% for RECaST-Surv, compared with 81.2\% for the RCT-only analysis. The estimated hazard ratio was 0.708 for RECaST-Surv and 0.680 for the RCT-only analysis. The MSE was slightly larger for RECaST-Surv, reflecting the additional population shift introduced by biased trial sampling. Nevertheless, the gain in power remained substantial.

Overall, the ALS emulation suggests that RECaST-Surv can deliver meaningful gains in power in a realistic trial setting. The random sampling setting shows the expected benefit of external borrowing when the trial and registry populations are well aligned. The biased sampling setting further suggests that the method can still maintain type I error control and improve power when the trial cohort differs from the external registry population, although some shift in the hazard ratio estimate may occur when the borrowed external information is less directly aligned with the emulated trial.





\section{Discussion and Conclusion} \label{sec:conclude}

In this paper, we proposed the RECaST-Surv design, a Bayesian transfer-learning framework for borrowing information from external data in randomized clinical trials with time-to-event endpoints. The method extends the original RECaST framework to survival outcomes and is particularly useful in trials with limited concurrent control information, especially unequal randomized trials where the concurrent control arm is relatively small, although the same framework can also be used in 1:1 randomized trials when external borrowing is beneficial. The basic idea is to use the external data to learn a source model, calibrate this model to the current trial control arm through the RECaST framework, and then generate counterfactual survival predictions for the treated patients. In this framework, the latent variable $\beta_i$ is treated as a calibration random effect that adjusts the source-model linear predictor to the target population, rather than as a conventional patient-specific regression coefficient. This leads to a natural way to estimate the treatment effect in the treated population.

A main feature of the proposed design is the bootstrap-based calibration step. Rather than relying only on prespecified simulation scenarios to study type I error, we use the observed control-arm data from the current trial to construct a study-specific rejection rule that reflects the randomization structure of the trial. We do not claim that this guarantees exact nominal type I error in finite samples. Instead, the goal is to provide a practical and data-driven way to account for type I error at the trial level using the currently observed data. In our simulation studies, the proposed design generally produced type I error rates close to the nominal level while yielding a clear gain in power over the no-borrowing analysis. In the real-data example, the method also showed a substantial gain in power, with only mild type I error inflation.

The proposed framework is also attractive in practice because, once an external source model has been appropriately constructed, only the fitted source-model parameters are required for the RECaST-Surv analysis. This is useful in settings where patient-level data cannot be easily shared across institutions because of privacy, administrative, or logistical constraints. This advantage does not remove the need for appropriate construction of the external source model, including harmonization of eligibility criteria, covariate and outcome definitions, and the time origin for survival follow-up. When multiple external sources are available, the proposed weighting scheme further allows the method to combine source-specific predictions while down-weighting sources that appear less compatible with the current trial.

There are several limitations. First, the current implementation relies on a parametric Weibull survival model, which may be restrictive when the underlying hazard structure is more complex. Second, although the calibration step helps address type I error in a study-specific way, some inflation may still occur in practice, especially when the external data differ materially from the current trial. Third, the treatment effect test in this paper is based on a one-sample log-rank-type procedure constructed from the predicted counterfactual survival curves. While this provides a practical working solution, other choices are possible, such as procedures based on restricted mean survival time. 
Finally, the quality of borrowing depends on the relevance of the external sources and on the adequacy of the source-model construction. An important consideration for registry or other observational external controls is the definition of baseline or time zero. Unlike an RCT, in which eligibility assessment and randomization provide a natural start of follow-up, observational patients may become eligible at different points in their disease course. If the source model is estimated using a time origin that is not appropriately aligned with the target trial, selection or immortal-time bias may be incorporated into the estimated source-model parameters. The current RECaST-Surv calibration can adapt the source model to the concurrent trial controls, but it is not designed to reconstruct the external risk sets or guarantee removal of bias caused by an inappropriate definition of time zero. 

There are several directions for future work. One is to extend the framework to settings with time-varying covariates or competing risks. Another is to study alternative test statistics and calibration rules within the same general design. It would also be useful to consider more flexible survival models within the RECaST framework, so that the borrowing procedure is less sensitive to parametric assumptions. Another direction is to integrate RECaST-Surv with target-trial or sequential-emulation strategies when observational external controls do not have a natural trial-aligned time zero. Under such an approach, the external data could be appropriately indexed and analyzed locally, while only the resulting source-model information would need to be transferred to the RECaST-Surv analysis. More generally, future extensions could incorporate richer source-level summaries, such as uncertainty in the estimated source parameters, baseline survival information, or estimates obtained under alternative index-date definitions, to evaluate and potentially calibrate source-target differences without requiring transfer of the complete patient-level external data. 

Overall, the proposed RECaST-Surv design provides a practical framework for borrowing external information in survival trials with limited concurrent controls. The method gives a clear gain in power in our studies, while keeping type I error inflation limited. As with other approaches using observational external controls, appropriate construction of the external source model, including alignment of eligibility, outcome definitions, and the time origin with the target trial, remains important for reliable application. 

\bibliographystyle{plainnat}
\bibliography{reference}

\newpage

\appendix

\section{}

\renewcommand\thefigure{\thesection.\arabic{figure}} 
\setcounter{figure}{0}

\renewcommand\thetable{\thesection.\arabic{table}} 
\setcounter{table}{0}

\renewcommand\theequation{\thesection.\arabic{equation}} 
\setcounter{equation}{0}

\subsection{MCMC Convergence Analysis}\label{subsec:mcmc_ana}

\begin{figure}[h]
  \centering
  \includegraphics[width=\textwidth]{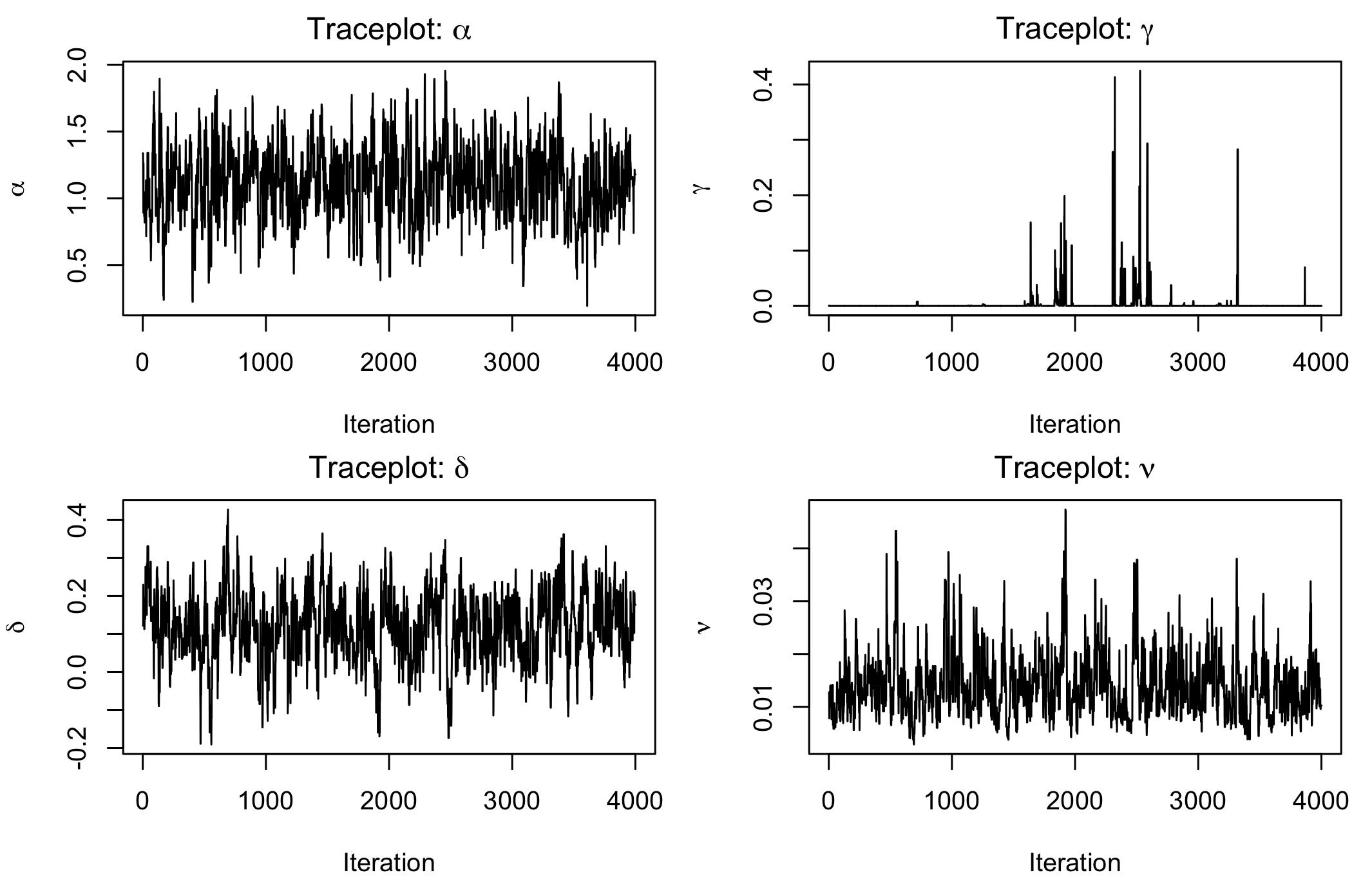}
  \caption{Representative trace plots for $(\alpha,\delta,\gamma,\nu)$ from a randomly selected congruent null-scenario dataset, showing reasonably stable behavior after burn-in. The chain for $\gamma$ is concentrated near zero with occasional spikes.}
  \label{fig:traceplot}
  
\end{figure}

Figure \ref{fig:traceplot} shows representative trace plots for $(\alpha,\delta,\gamma,\nu)$ under a randomly selected congruent null scenario dataset suggest no obvious long-term drift after burn-in. The chains for $\alpha$, $\delta$, and $\nu$ show reasonably stable mixing, while the chain for $\gamma$ is concentrated near zero with occasional spikes, which is consistent with a posterior distribution that places most of its mass on small calibration-scale values.

\subsection{Expected Number of Events in Simulation}\label{subsec:exp_events}

Table \ref{tab:expected_events} shows the expected number of events in the current randomized trial across the two sample sizes and the three allocation ratios.

\begin{table}[ht]
\centering
\caption{Expected number of events in the treated and control arms under different trial sample sizes, randomization ratios, and treatment effects. Results are based on the simulation summaries.}
\label{tab:expected_events}
\small
\begin{tabular}{cc | cc | cc | cc}
\toprule
 &  & \multicolumn{2}{c|}{$\text{HR} = 1$} 
    & \multicolumn{2}{c|}{$\text{HR} = 0.779$} 
    & \multicolumn{2}{c}{$\text{HR} = 0.607$} \\
\cmidrule(lr){3-4} \cmidrule(lr){5-6} \cmidrule(lr){7-8}
$N$ & $r$ 
& Treated & Control 
& Treated & Control 
& Treated & Control \\
\midrule
200 & 1 & 67.9 & 68.1 & 60.4 & 67.7 & 52.6 & 67.9 \\
200 & 2 & 90.5 & 45.3 & 80.4 & 45.1 & 70.5 & 45.3 \\
200 & 3 & 101.8 & 33.8 & 90.7 & 34.2 & 79.3 & 33.8 \\
300 & 1 & 102.0 & 102.2 & 91.0 & 101.4 & 79.2 & 101.8 \\
300 & 2 & 136.1 & 68.0 & 120.8 & 68.4 & 105.6 & 68.2 \\
300 & 3 & 152.9 & 51.0 & 136.0 & 50.9 & 118.4 & 51.0 \\
\bottomrule
\end{tabular}
\end{table}

\subsection{Sensitivity Analysis}\label{subsec:sens_ana}

\paragraph{Simulation Setup:}
For all sensitivity scenarios, we fixed the current trial sample size at $N=300$ with a $2:1$ randomization ratio. Survival outcomes were generated from the same data-generating mechanism as in Section 4.1 of the main manuscript, using a Weibull proportional hazards model with shape parameter $\nu=1.2$ and scale parameter $\lambda=0.01$.

The stress scenarios were defined as follows:
\begin{itemize}
    \item \textbf{Scenario Set A: external sample size and coefficient magnitude.} We varied the sample sizes of the three external sources, $(n_1,n_2,n_3)$, as well as the magnitude of their regression coefficients relative to those in the current trial.
    \begin{itemize}
        \item Cases 1--3: mixed sample sizes. In these cases, the external sources had different sample sizes, and their coefficient vectors were either close to or different from that of the current trial. Specifically, we let $\bm{\theta}_0=(0.20,0.30,0.40)^T$. We then set $\bm{\theta}_1=\bm{\theta}_0$, $\bm{\theta}_2=(0.25,0.35,0.45)^T$, and $\bm{\theta}_3=(0.15,0.25,0.35)^T$. We fixed $n_1=50$ in all three cases, and let $n_2$ and $n_3$ vary in $\{200,350,500\}$.
        \item Cases 4--5: systematic scaling. In these cases, all external coefficient vectors were set to be systematically larger or systematically smaller than the trial coefficient vector. The shifts were approximately $+0.05$ to $+0.15$ in one case and $-0.05$ to $-0.15$ in the other.
    \end{itemize}

    \item \textbf{Scenario Set B: covariate correlation.} We introduced correlation between covariates $X_1$ and $X_2$ with $\rho \in \{-0.5,0.5\}$. This correlation structure was imposed either symmetrically on both the trial and external data, or asymmetrically across the two.

    \item \textbf{Scenario Set C: parameter corruption.} We perturbed the estimated source coefficients $\hat{\bm{\theta}}_j, \, j > 0,$ to mimic unreliable reporting from the external data. This was done either by adding a fixed bias ($\pm 0.5$) or by adding random noise with standard deviation 1.0.

    \item \textbf{Scenario Set D: biased sampling (covariate shift).} We used rejection sampling to create disjoint populations between the current trial and the external data. For a selected covariate $X_1$, the trial data were restricted to one region of the covariate space, for example $X_1 \in [0,\infty)$, while the external data were restricted to a different region, for example $X_1 \in (-\infty,0)$.
\end{itemize}

For each sensitivity setting, we generated 1{,}000 simulated datasets using the same trial configuration as in Section 4.1 of the main manuscript. The results are summarized in Tables \ref{tab:sensitivity_results1} and \ref{tab:sensitivity_results2}.

\begin{table}[ht]
\centering
\caption{Sensitivity Analysis Results across Scenario sets A and B.}
\label{tab:sensitivity_results1}
\resizebox{\textwidth}{!}{%
\begin{tabular}{c c p{5cm} | c c | c c}
\toprule
\multirow{2}{*}{\textbf{Scenario}} & \multirow{2}{*}{\textbf{Case}} & \multirow{2}{*}{\textbf{Description}} & \multicolumn{2}{c}{\textbf{Null (HR = 1.000)}} & \multicolumn{2}{c}{\textbf{Alt. (HR = 0.607)}} \\
\cmidrule(lr){4-5} \cmidrule(lr){6-7}
 & & & \textbf{Est.} & \textbf{TIER} & \textbf{Est.} & \textbf{Power} \\
\midrule
\multirow{5}{*}{A} 
 & 1 & $n_2 = 500$ and $n_3 = 200$ & 1.026 & 0.054 & 0.658 & 0.903 \\
 \cline{2-7}
 & 2 & $n_2 = 350$ and $n_3 = 350$ & 1.013 & 0.051 & 0.648 & 0.929 \\
 \cline{2-7}
 & 3 & $n_2 = 200$ and $n_3 = 500$ & 1.014 & 0.049 & 0.648 & 0.924 \\
 \cline{2-7}
 & 4 & All external sources coefficents are larger than trial & 1.020 & 0.038 & 0.654 & 0.921 \\
 \cline{2-7}
 & 5 & All external sources coefficents are smaller than trial & 1.014 & 0.058 & 0.647 & 0.920 \\
\midrule
\multirow{10}{*}{B} 
 & 1 & $\rho = -0.5$ in trial patients & 1.023 & 0.054 & 0.649 & 0.909 \\
 \cline{2-7}
 & 2 & $\rho = 0.5$ in trial patients & 1.022 & 0.053 & 0.660 & 0.902 \\
 \cline{2-7}
 & 3 & $\rho = -0.5$ in all external sources & 1.018 & 0.055 & 0.651 & 0.910 \\
 \cline{2-7}
 & 4 & $\rho = 0.5$ in all external sources & 1.019 & 0.059 & 0.651 & 0.927 \\
 \cline{2-7}
 & 5 & $\rho = 0.5$ in both trial and all external sources patients & 1.024 & 0.058 & 0.662 & 0.909 \\
 \cline{2-7}
 & 6 & $\rho = 0.5$ in trial patients and $\rho = -0.5$ in all external sources & 1.019 & 0.049 & 0.659 & 0.907 \\
 \cline{2-7}
 & 7 & $\rho = -0.5$ in trial patients and $\rho = 0.5$ in all external sources & 1.020 & 0.041 & 0.646 & 0.931 \\
 \cline{2-7}
 & 8 & $\rho = -0.5$ in both trial and all external sources & 1.021 & 0.044 & 0.648 & 0.928 \\
\bottomrule
\end{tabular}%
}
\end{table}

\begin{table}[ht]
\centering
\caption{Sensitivity Analysis Results across Scenario sets C and D.}
\label{tab:sensitivity_results2}
\resizebox{\textwidth}{!}{%
\begin{tabular}{c c p{5cm} | c c | c c}
\toprule
\multirow{2}{*}{\textbf{Scenario}} & \multirow{2}{*}{\textbf{Case}} & \multirow{2}{*}{\textbf{Description}} & \multicolumn{2}{c}{\textbf{Null (HR = 1.000)}} & \multicolumn{2}{c}{\textbf{Alt. (HR = 0.607)}} \\
\cmidrule(lr){4-5} \cmidrule(lr){6-7}
 & & & \textbf{Est.} & \textbf{TIER} & \textbf{Est.} & \textbf{Power} \\
\midrule
\multirow{11}{*}{C} 
 & 1 & Add random noise to external source 1. & 1.019 & 0.063 & 0.651 & 0.903 \\
 \cline{2-7}
 & 2 & Add +0.5 bias to external source 2. & 1.017 & 0.052 & 0.651 & 0.930 \\
 \cline{2-7}
 & 3 & Add -0.5 bias to external source 2 & 1.021 & 0.051 & 0.653 & 0.917 \\
 \cline{2-7}
 & 4 & Add random noise and +0.5 bias to external source 3. & 1.021 & 0.056 & 0.651 & 0.917 \\
 \cline{2-7}
 & 5 & Add random noise and -0.5 bias to external source 3. & 1.014 & 0.047 & 0.648 & 0.927 \\
 \cline{2-7}
 & 6 & Add random noise to external source 2. & 1.023 & 0.061 & 0.654 & 0.906 \\
 \cline{2-7}
 & 7 & Add random noise to external source 3. & 1.018 & 0.058 & 0.649 & 0.928 \\
 \cline{2-7}
 & 8 & Add random noise to external sources 1 and 2. & 1.021 & 0.053 & 0.653 & 0.898 \\
 \cline{2-7}
 & 9 & Add random noise to external sources 2 and 3. & 1.012 & 0.051 & 0.646 & 0.918 \\
\cline{2-7}
 & 10 & Add random noise to external sources 1 and 3. & 1.026 & 0.046 & 0.655 & 0.912 \\
 \cline{2-7}
 & 11 & Add random noise to all external sources. & 1.017 & 0.070 & 0.649 & 0.901 \\
\midrule
\multirow{5}{*}{D} 
 & 1 & The trial paitents restricted to $[0, \infty)$. & 1.022 & 0.049 & 0.654 & 0.928 \\
 \cline{2-7}
 & 2 & The 1st external source paitents restricted to $[0, \infty)$. & 1.021 & 0.060 & 0.653 & 0.909 \\
 \cline{2-7}
 & 3 & All external sources paitents restricted to $[0, \infty)$ & 1.025 & 0.049 & 0.656 & 0.908 \\
 \cline{2-7}
 & 4 &  Trial paitents $[0, \infty)$ and source 1 paitents $(-\infty, 0]$. & 1.021 & 0.060 & 0.652 & 0.925 \\
 \cline{2-7}
 & 5 & Trial paitents $[0, \infty)$ and all sources paitents $(-\infty, 0]$. & 1.013 & 0.052 & 0.649 & 0.926 \\
\bottomrule
\end{tabular}%
}
\end{table}

\end{document}